\pdfoutput=1

\documentclass[english]{article}
\usepackage{times}
\usepackage[labelfont=bf]{caption}
\usepackage[T1]{fontenc}
\usepackage[latin1]{inputenc}
\usepackage{geometry}
\usepackage{rotating}
\usepackage{graphicx}
\usepackage{setspace}
\usepackage[authoryear]{natbib}

\usepackage{amssymb}
\usepackage{amsmath}

\makeatletter

\usepackage{babel}
\makeatother
\begin{document}

\title{Identifying Signatures of Selection \\ in Genetic Time Series \\
}

\author{~\\
\\
\\
\\
 \\
Alison F. Feder$^{1,*}$, Sergey Kryazhimskiy$^{2,*}$, Joshua B. Plotkin$^1$}

    %[AF:EDIT] Added middle initial 

\date{~ }

\maketitle
~\\
\\
\\
\\
\\
\\
\\
\\
\\
\\
${}^1$Department of Biology, University of Pennsylvania, Philadelphia, PA

\noindent ${}^2$Department of Organismic and Evolutionary Biology and FAS Center for Systems Biology, Harvard University, Cambridge, MA

\noindent ${}^*$Equal contribution

\clearpage

Running Head: Selection in genetic time series \\

Keywords: \\

Corresponding author:

Joshua B. Plotkin

Department of Biology

University of Pennsylvania

Philadelphia

PA 19104

USA

Phone 

Fax 

Email: jplotkin@sas.upenn.edu

\clearpage

\begin{abstract} {\normalsize Both genetic drift and natural selection cause
the frequencies of alleles in a population to vary over time. Discriminating between
these two evolutionary forces, based on a time series of samples from a
population, remains an outstanding problem with increasing relevance to modern
data sets.
 Even in the idealized situation when
the sampled locus is independent of all other loci this problem is
difficult to solve, especially when
the size of the population from which the samples are drawn is unknown.  A standard
$\chi^2$-based likelihood ratio test was previously proposed to address this
problem.  Here we show that the $\chi^2$ test of selection substantially underestimates
the probability of Type I error, leading to more false positives than indicated by
its $P$-value, especially at stringent $P$-values.  We introduce two methods
to correct this bias.  The empirical likelihood ratio test (ELRT) rejects
neutrality when the likelihood ratio statistic falls in the tail of the
empirical distribution obtained under the most likely neutral population size.
The frequency increment test (FIT) rejects neutrality if the
distribution of normalized allele frequency increments exhibits a mean that
deviates significantly from zero.
 We characterize the statistical power of these two tests for selection, and we apply them to
three experimental data sets.  We demonstrate that both ELRT
and FIT have power to detect selection in practical parameter regimes, such as
those encountered in microbial evolution experiments.  Our analysis
applies to a single diallelic locus, assumed independent of all other loci,
which is most relevant to full-genome selection scans in sexual organisms, and also to evolution experiments in asexual organisms as long as clonal interference is weak. Different techniques will be required to detect selection in time series of co-segregating linked loci.  }{\normalsize \par} \end{abstract}
\clearpage

\section{Introduction}

Population geneticists typically seek to understand the forces responsible for patterns observed in contemporaneous samples of genetic data, such as
the nucleotide differences fixed between species, polymorphisms within populations, and the structure of linkage disequilibrium.  Recently, however,
there has been a rapid increase in the availability of dynamic data, where the frequencies of segregating alleles in an evolving population are
monitored through time, both in laboratory experiments \citep{Hegreness2006,BollbackHuelsenbeck2007,Barrick2009,LangBotsteinDesai2011,Orozco2012,LangRice2013} and in natural populations
\citep{Barrett2008,Reid2011,DenefBanfield2012,Winters2012,Daniels2013,Maldarelli2013,Pennings2013}.  One important question is whether the changes in
allele frequencies observed in such data are the result of natural selection or are simply consequences of genetic drift or sampling noise.
% If natural selection is implicated, we would also like to know how strong the selective forces are, and the relevant effective population sizes.
In principle, it seems that dynamic data should provide researchers with more power to detect and quantify selective forces while avoiding the
assumptions of stationarity that are required for many inference techniques based on static samples
\citep{SawyerHartl1992,DesaiPlotkin2008,Boyko2008}.  Nonetheless, the behavior and power of inference techniques based on time series data have not
been thoroughly investigated.

There is a well-developed literature on inferring population sizes from genetic time-series data assuming neutrality
\citep{Pollak1983,Waples1989,WilliamsonSlatkin1999,Wang2001} and a rapidly growing literature on inferring natural selection from such time series
\citep{Bollback2008,IllingworthMustonen2011,Illingworth2012,Malaspinas2012,Mathieson2013}.  However, even the simplest case -- the dynamics of two
alternative alleles at a single genetic locus independent of all other loci -- presents a number of statistical challenges that have not been
resolved.  The main complication arises when the actual size of the population from which the serial samples are drawn is unknown.  In this case,
large changes in the frequency of an allele might indicate either that the allele is under selection, or that the population size is small and genetic
drift is strong.  To favor one alternative over the other \citet{Bollback2008} proposed to fit two nested Wright-Fisher models to time-series data at
a single locus (one model with selection, and one without) and reject the neutral model using the $\chi^2$ distribution for the likelihood ratio
statistic.  Such an approach is generally the most powerful and unbiased, at least for large data sets.  Nonetheless, here we show that in practice the
actual frequency of false positives under this approach can vastly exceed the nominal $P$-value obtained from the $\chi^2$ distribution -- and
especially so at more stringent $P$-value cutoffs.  Since the $\chi^2$ distribution does not provide an accurate representation of the false positive
rate, this approach cannot be used to draw sound statistical conclusions about selection from such time series.  The underlying reason for this
problem is that the likelihood ratio statistic is $\chi^2$-distributed only asymptotically, and convergence to this distribution is slow
\citep{Wilks1938}.  In most practical applications, such as when sampling from natural populations
\citep{Reid2011,DenefBanfield2012,Winters2012,Daniels2013,Maldarelli2013,Pennings2013} or competing two microbial strains
\citep{Lenski1991,BollbackHuelsenbeck2007,LangRice2013}, the number of sampled time points is typically small (fewer than 10) and the distribution of
the likelihood ratio statistic is far from $\chi^2$ under neutrality, leading to more false positives then expected.

We propose two solutions to fix this problem, providing unbiased tests for natural selection in time-series data sampled at a single genetic locus.
First, we develop an algorithm for computing the exact distribution of the likelihood ratio statistic under neutrality.  Although feasible in many
regimes, this direct approach suffers from several complications which we discuss below.  We also propose an alternative, computationally efficient,
albeit approximate, statistical method for rejecting the neutral model.
%  We are able to estimate the true false-discovery rate for both of our methods. And we show that both methods have power to detect selection from
%  time-series of allele-frequencies.
Our approach builds directly on the work of \citet{Bollback2008}, and it is likewise limited to studying time series of allele frequencies at a single
locus under genic selection, assuming independence from all other loci. The more complicated problem of detecting selection from genomic time series
of many linked loci has received attention elsewhere \citep{IllingworthMustonen2011,Illingworth2012}, and the problems identified here likely apply to
those situations as well.

We start our presentation by introducing a likelihood framework for time-series data at a single genetic locus.
%, and we then derive approximate expressions for these likelihoods under the Gaussian approximation to the Wright-Fisher model.
We then demonstrate that the $P$-value given by the $\chi^2$ distribution for the likelihood ratio statistic underestimates the actual false-discovery
rate.  Next, we introduce two methods to correct this bias, and we verify that they are virtually unbiased for large sample sizes and conservative for
small sample sizes. We quantify the power of these two tests for selection in different parameter regimes, considering also noise in the measurements
of allele frequencies.  Finally, we apply our methods to three experimental data sets and demonstrate that the tests behave as expected in practical
situations.

    %[AF:EDIT] measurments -> measurements 
    
\section{Materials and Methods}

\subsection{Approximate expression of the transition probability for the Moran process}

Calculating the likelihood of an allele-frequency time series requires knowing the
transition probability $P_s(x, t|x^\prime, t^\prime)$ that the frequency of the
observed allele in the population at time $t$ is $x$, given that it was $x^\prime$
at some previous time $t^\prime$.  The sub-script $s$ indicates that this
probability depends on the selection coefficient of the allele.  In general, it
will also depend on the population size $N$ and maybe on other parameters.  Under
most population-genetic models no exact analytical expressions for the transition
probability $P_s(x, t|x^\prime, t^\prime)$ are available for arbitrary $x,
x^\prime, t, t^\prime$, and $s$.  The standard approximation to the discrete
Wright-Fisher and Moran models is the diffusion approximation of Kimura and others
\citep{Ewens}.  Although considerably simpler than the discrete models, the
diffusion equation is still difficult to solve exactly and efficiently in a
general case.  Although some numerical methods are available
\citep{Kimura1955PNAS,Kimura1955CSH,Evans2007,Bollback2008,SongSteinrucken2012},
they are often cumbersome to implement or computationally intensive.

Therefore, we will use a Gaussian approximation to the Wright-Fisher process,
which is less accurate than the diffusion approximation but allows us to obtain a
simple analytical expression for the transition probability, which can be computed
efficiently and is quite accurate provided the allele has not been lost or fixed
during the period of observation. We emphasize that our two tests for
selection  proposed below do not intrinsically depend upon this Gaussian approximation (that is,
they could in principal be implemented using the full Wright-Fisher model or
the Kimura diffusion), but we nonetheless rely on this approximation for efficiency's
sake. Moreover, as we will discuss below, there is little additional power to be gained
by considering time-series that exhibit many sampled time points with fixed alleles, provided that sampling noise is small.

 %[AF:EDIT] timepoints -> time points
We describe the Gaussian approximation in detail in the Appendix and summarize it here.
Briefly, if the timescale of observation is short compared to $N$ in case of a
  neutral allele or to $1/s $ in case of a positively selected allele, i.e., if
  absorption events can be neglected, the Moran process can be approximated by a sum of a deterministic process $g$ and a Gaussian noise process $Z$ \citep{Pollett1990}.
  In the absence of genetic drift, i.e., when $N \to \infty$, the allele frequency $X$ behaves deterministically, $X \to g(t, x_0)$, where $g$ satisfies the logistic equation
  \begin{eqnarray}
    \dot g & = & s g ( 1- g), \label{eq: dg/dt} \\
    g(0, x_0) & = & x_0, \label{eq: g(0)}
  \end{eqnarray}
whose solution is
\begin{equation}
  g(t, x_0) = x_0 \left(x_0 + (1-x_0)e^{-st}\right)^{-1}. \label{eq:g(t)}  
\end{equation}
Here $x_0$ is the initial deterministic allele frequency.
  When $N<\infty$, genetic drift perturbs the allele frequency $X$ from its deterministic value and so $X(t) = g(t, x_0) + Z(t)$ where $Z(t)$ is the
  noise process.
  Then for any two time points $t^\prime \geq 0$ and $t > t^\prime$, the transition probability is approximated by
  \begin{eqnarray}
    P_s(x,t|x', t') & \approx & \sqrt{\frac{N}{2 \pi \sigma^2\left(\Delta t, g^\prime\right)}} \exp \left\{
      -N \frac{ \big( x -  g  - \left(x' - g^\prime\right)M\left(\Delta t, g^\prime\right) \big)^2  }{2 \sigma^2\left(\Delta t, g^\prime\right)}
    \right\},      \label{eq:P_approx}
  \end{eqnarray}
where
  \begin{eqnarray}
    M(\Delta t, \xi) & = & e^{-s\Delta t} \left(\xi + (1 - \xi) e^{-s\Delta t} \right)^{-2},  \label{eq:M(t)} \\
    \sigma^2(\Delta t, \xi) & = & M^2(\Delta t, \xi) (2 + s) \xi (1-\xi)s^{-1}  \nonumber \\
    & \times & \Big[
      2 \xi (1-\xi)s\Delta t + \xi^2e^{s \Delta t}-(1-\xi)^2e^{-s \Delta t} + (1-\xi)^2 -\xi^2
    \Big]    \label{eq:var(t)}
  \end{eqnarray}
and we used shorthands $g \equiv g(t, x_0) $, $g^\prime \equiv g(t^\prime, x_0) $, $\Delta t = t - t^\prime$.
  Under the neutral null hypothesis (i.e., when $s = 0$), the transition probability simplifies to
  \begin{eqnarray}
    P_0(x,t|x', t') & \approx & \sqrt{\frac{N}{2 \pi \sigma_n^2(\Delta t, x_0)}} \exp \left\{
      -N \frac{ \left( x - x' \right)^2  }{2 \sigma_n^2(\Delta t, x_0)}
    \right\},      \label{eq:P_approx neut}
  \end{eqnarray}
with
  \begin{equation}
    \sigma_n^2(\Delta t, x_0) = 2x_0(1-x_0) \Delta t.  \label{eq:var(t) neut} 
  \end{equation}
  Note that functions \eqref{eq:g(t)}--\eqref{eq:var(t) neut} depend on parameters $N$, $s$ and on the nuisance parameter  $x_0$ that in principle can be estimated along with $N$ and $s$.
  However, for the sake of reducing the number of fitted parameters, we fix $x_0 $ to be equal to the observed allele frequency at time zero, $x_0 \equiv \nu_0 $.

  We assume here that time is measured in generations.
  If time is measured in physical units, equations (\ref{eq:g(t)}), (\ref{eq:M(t)}) still hold, with rescaled parameters $N \to N \tau$, and $s \to s/\tau$, where $\tau$ is the generation time; equation (\ref{eq:var(t)}) does not hold exactly because of the term $2+s$, but holds approximately as long as $s \ll 1$, which is true in most cases.
  Thus, equations (\ref{eq:P_approx}), (\ref{eq:P_approx neut}) can still be used.

\subsection{Implementation}

  In the Results and Discussion section, we obtain the expression for the likelihood $L(\mathrm{Data}; N, s)$ of allele-frequency data as a function of two parameters, $N$ and $s$.
  We estimate these parameters by maximizing this likelihood expression.
  First, consider the case when the allele frequency is measured at only two time points $t_0$ and $t_1$ with the corresponding frequencies being $\nu_0$ and $\nu_1$.
  Then the likelihood expression (\ref{eq:llh nu}) with the Gaussian approximation (\ref{eq:P_approx}) becomes
  \begin{displaymath}
    L(\textrm{Data}; N, s)
    = \sqrt{\frac{N}{2 \pi \sigma^2\left(\Delta t, \nu_0 \right)}} \exp \left\{
      -N \frac{ \big( \nu_1 -  g(t_1, \nu_0) \big)^2  }{2 \sigma^2\left(\Delta t, \nu_0 \right)}
    \right\}, 
  \end{displaymath}
which is maximized at $ \hat N = \infty $ and
\begin{displaymath}
  \hat s = \frac{1}{t_1-t_0}\ln\left( \frac{\nu_1}{1-\nu_1}\frac{1-\nu_0}{\nu_0} \right).
\end{displaymath}
  In this case, the Gaussian likelihood function collapses to a delta-function centered at $\hat s$ so that $L(\textrm{Data}; \hat N, \hat s) = \infty $. 
  In other words, with two data points there is enough information to estimate only the selection coefficient but not the population size.
  Thus, the likelihood ratio approach can only be applied to three or more sampled time points, in which case we find the maximum likelihood parameter values using the Nelder-Mead simplex method \citep{NelderMead1965} implemented in the Gnu Scientific Library (GSL) package.
  We limit the search to the interval $[-2, 2]$ for $s$ (although in practice $|s| \ll 1$) and to the interval
  $[10^{-1}, 10^8]$ for $N$, and we allow a maximum of $3 \times 10^4$ function evaluations.

  Even though the frequency increment test described below does not rely on the calculation of $\hat N $ and $\hat s$, it too can only be applied when three or more sampled time points are available, for the same conceptual reason as described above.
  Mathematically, when only one frequency increment is observed, the variance of the distribution of increments cannot be estimated and the $t$-statistic cannot be computed.

% \subsection{Competition assay}
%   The details of the evolution experiment

%   The competition experiment consists of mixing the evolved population with a fluorescently marked reference strain, propagating them together for a relatively small number of generations (e.g., 20) and measuring the change in frequency of the evolved population relative to the reference strain \citep{Lenski1991,Kryazhimskiy2012}.

\section{Results and Discussion}

  We consider the problem of determining whether selection has played a role in
  shaping the fluctuations in the observed frequencies of an allele in a
  population sampled over time.
  Suppose that at each time point $t_i$ ($0 < t_1 < \dots < t_L$) we sample a diallelic locus in $n_i$ individuals from a given population of an unknown size $N$ and observe that $b_i$ individuals carry allele $A$ and $1 - b_i$ individuals carry allele $a$.
% 
  %[AF, SK:EDIT]
  %OLD:   Suppose that at time points $0 < t_1 < \dots < t_L$ we sample $n_0, n_1, \dots, n_L$ individuals from a given population of an unknown size $N$ and observe that $b_i$ of them carry allele $A$ at the locus of interest.
 % 
  Thus, we observe sampled allele frequencies $\nu_0 = b_0/n_0, \nu_1 = b_1/n_1, \dots, \nu_L = b_L/n_L$.
  We ask whether genetic drift and sampling noise alone are sufficient to explain
  the fluctuations in the sampled allele frequencies, or whether these frequency
  changes implicate the action of natural selection at either the specified locus or another completely linked locus.
%  
    %[AF:EDIT]
    %OLD: whether these frequency  chagnes implicate the action of natural selection at either the specified locus  another completley linked locus.
    %NEW: whether these frequency changes implicate the action of natural selection at either the specified locus or another completely linked locus.
% 
 Initially, we treat this problem while neglecting sampling noise.
 That is, we initially assume that $n_i \gg 1$, $1 \ll b_i \ll n_i$ for all $i$,
 so that the sampled allele frequencies $\nu_i $ accurately represent the actual
 frequencies in the entire population.
  We later investigate how sampling noise affects our conclusions.  

We approach the problem using the standard likelihood ratio test.  Following
\citep{Bollback2008}, we consider a pair of nested hypotheses.  Under the neutral
null hypothesis, changes in the allele frequency are caused only by genetic drift,
i.e., the selection coefficient $s$ of allele $A$ is assumed to be zero.  Under
the alternative hypothesis there is no restriction on $ s $.  In both cases, allele $a$ is not under selection.
%
    %[AF:EDIT]
    %Inserted: In both cases, allele $a$ is not under selection.
We calculate the
likelihoods of the allele-frequency time series under each of these hypotheses,
compute the likelihood ratio statistic (LRS), and reject neutrality if the LRS falls
in the tail of the $\chi^2$ distribution with one degree of freedom.  Because the LRS need 
%
    %[AF:EDIT]
    %OLD: reject neutrality if thre LRS falls in the tail of the $\chi^2$ distribution with one degree of freedom.  Because thre LRS need 
   %NEW: reject neutrality if the LRS falls in the tail of the $\chi^2$ distribution with one degree of freedom.  Because the LRS need 
%  
not be
$\chi^2$-distributed when the number of data points is small, we first report
comparisons between the $\chi^2$ distribution and the true distribution of the LRS, for a
range of sample sizes.  To do this, we simulate samples from the neutral Wright-Fisher
process and report whether the probability of Type I error in
the $\chi^2$ test is accurately predicted by the associated $\chi^2$ $P$-value.

\subsection{Likelihood of time-series data and the likelihood ratio statistic}
  Under standard single-locus population-genetic models, the dynamics of an allele with selection coefficient $s$ in a population are described by a Markov process that specifies the transition probability $P_s(x, t|x^\prime, t^\prime)$ that the allele frequency is $x$ at time $t$, given that it was $x^\prime$ at some previous time $t^\prime$.
  In addition to the selection coefficient $s$, this transition probability depends also on the population size $N$ and possibly on other nuisance parameters \citep{Ewens}.
  Ignoring sampling noise, the likelihood of observing allele frequencies $\nu_0, \nu_1, \dots, \nu_L$ at times $0, t_1, \dots, t_L $ is
  \begin{equation}
    \label{eq:llh nu}
    L(\mathrm{Data}; N, s) = U(\nu_0) \prod_{i=1}^L P_s(\nu_i, t_i | \nu_{i-1}, t_{i-1} ),
  \end{equation}
and, under the neutral null hypothesis,
  \begin{equation}
    \label{eq:llh nu neut}
    L(\mathrm{Data}; N, 0) = U(\nu_0) \prod_{i=1}^L P_0(\nu_i, t_i | \nu_{i-1}, t_{i-1} ).
  \end{equation}
  Here $U(x)$ denote the probability of observing allele frequency $x$ at time point 0
  which for simplicity we set to be uniform on the interval $(0,1)$, i.e., $U(x)
  \equiv 1$.
%, reflecting our lack of prior expectation for $\nu_0$. If the prior distribution of $\nu_0$ is known, it can be trivially substituted into equations (\ref{eq:llh nu}),~(\ref{eq:llh nu neut}).

  Computing likelihoods \eqref{eq:llh nu},
  \eqref{eq:llh nu neut} is non-trivial even for the standard Wright-Fisher
  process, because no exact analytical expression for the transition probability
  $P_s(x, t|x^\prime, t^\prime)$ exists, and approximate numerical procedures,
  based on the diffusion equation
  \citep{Kimura1955PNAS,Kimura1955CSH,Evans2007,Bollback2008,SongSteinrucken2012}
  are difficult to implement or computationally intensive.
  Since our investigation requires us to evaluate the likelihood function millions
  of times, we desire a fast algorithm for evaluating expressions \eqref{eq:llh nu}, \eqref{eq:llh nu neut}.
  Therefore, we choose to compute these likelihoods using analytical expressions obtained under the Gaussian approximation of the Wright-Fisher process, as is described in Materials and Methods and in the Appendix.
  Because the Gaussian approximation is accurate only when the allele frequency is far from 0 or 1, our results are restricted to time-series data that lack absorptions events.
%  However, we expect our conclusions to hold more generally, in cases when time series contain absorption events, when more sophisticated or even exact methods of computing likelihoods \eqref{eq:llh nu} and \eqref{eq:llh nu neut} are used.

  Given an algorithm for computing expressions (\ref{eq:llh nu}), (\ref{eq:llh nu
  neut}), we find the parameter values $\hat N$ and $\hat s$ that maximize the
  likelihood function (\ref{eq:llh nu}) and the value $\check N$ that maximizes
  the likelihood function (\ref{eq:llh nu neut}), and we compute the ratio,
  \begin{equation}
    R(\mathrm{Data}) = 2\log\left( \frac{L(\mathrm{Data}; \hat N, \hat s)}{L(\mathrm{Data}; \check N, 0)} \right).
    \label{eq:lrs}
  \end{equation}
  Note that the likelihood ratio statistic can be obtained only if the number of
  sampled time points is three or more, as explained in Materials and Methods.

  If our null hypothesis were {\em simple, }i.e., if the null distributions of the observed random variables did not depend on any free parameters, the Neyman-Pearson lemma would guarantee that the LRS defines the most powerful test of a given size for rejecting such null hypothesis \citep[Chapter 20]{StuartOrdArnold}.
  In other words, the Neyman-Pearson lemma instructs us to reject the null hypothesis whenever $R(\mathrm{Data}) > \varkappa_\alpha $ choosing $\varkappa_\alpha $ so that the probability of a Type I error is $\alpha$.
  This test is guaranteed to have the lowest probability of Type II error among all tests that have the same probability of Type I error, $\alpha$.

  In our case, however, the null hypotheses is {\em composite, }i.e., the
  distributions of allele frequencies depend on a parameter, $N$, whose value is unknown.
  This implies that the distribution of LRS under the null hypothesis is unspecified.
  Thus, not only is the likelihood ratio test not guaranteed to be the most powerful, there is no general way of determining the critical regions for the LRS distribution.
  The standard way to circumvent the latter problem is to use the asymptotic distribution for the LRS.
  When the number of data points approaches infinity, the LRS distribution converges
  to the $\chi^2$ distribution (in this case, with one degree of freedom), under appropriate regularity assumptions \citep{Wilks1938}.
  This approach has been previously used in the context of allelic time series by \citet{Bollback2008}.
  It is worth noting that, although the allele frequencies sampled at successive time points are not independent, the allele frequencies at successive time points {\em conditioned }on the frequencies at preceding time points are independent (this fact is reflected in expressions (\ref{eq:llh nu}),~(\ref{eq:llh nu neut})), and so the classical convergence results for LRS still hold.

  Although the LRS is guaranteed to be asymptotically $\chi^2$ distributed, the rate of convergence to this distribution is $O\left(1/\sqrt L\right)$ where $L$ is the number of sampled time points \citep{Wilks1938}.
  Therefore, we will characterize how well the $\chi^2$ distribution approximates the true distribution of LRS when the number of data points is finite.
  This question is important because the use of an incorrect null distribution
  can result in a test that underestimates the fraction of Type I errors, and thus
  erroneously rejects the null hypothesis more often than indicated by its $P$-value.

\subsection{Likelihood ratio statistic is not $\chi^2$ distributed for finite data }

  With this goal in mind, we simulated the neutral two-allele Wright-Fisher model
  with population size $N$, without mutation, with allele $A$ initiated at either
  10\%, 20\%, 30\%, 40\%, or 50\% of the population.
  We recorded the frequency of allele $A$ every generation, for $T$ generations.
  To ensure that absorption events are rare within the sampling period we set $T \leq N/10$.
  We then produced a data set consisting of these frequencies sampled every $\Delta$ generations.
  We sampled a total of $L+1$ time points, so that $\Delta = T/L$.
  For each population size $N$, we simulated $10^6$ allele-frequency trajectories,
  sampled allele frequencies from these trajectories using various combinations of
  $T$ and $L$, and computed the LRS for each of the sampled time series.
  Thus, for each combination of $N$, $L$ and $T$ we obtained the true distribution of LRS under the neutral null hypothesis.
  We compared this distribution with the $\chi^2$ distribution with 1 df in two ways.
  First, we calculated the probabilities for the LRS to fall into each of the 20 vigintiles (quantiles of size 0.05) of the $\chi^2$ distribution.
  Second, we computed the probability of Type I error of the $\chi^2$-based test for a range of nominal $P$-values $\alpha$.

\begin{figure}[h]
\begin{center}
\includegraphics{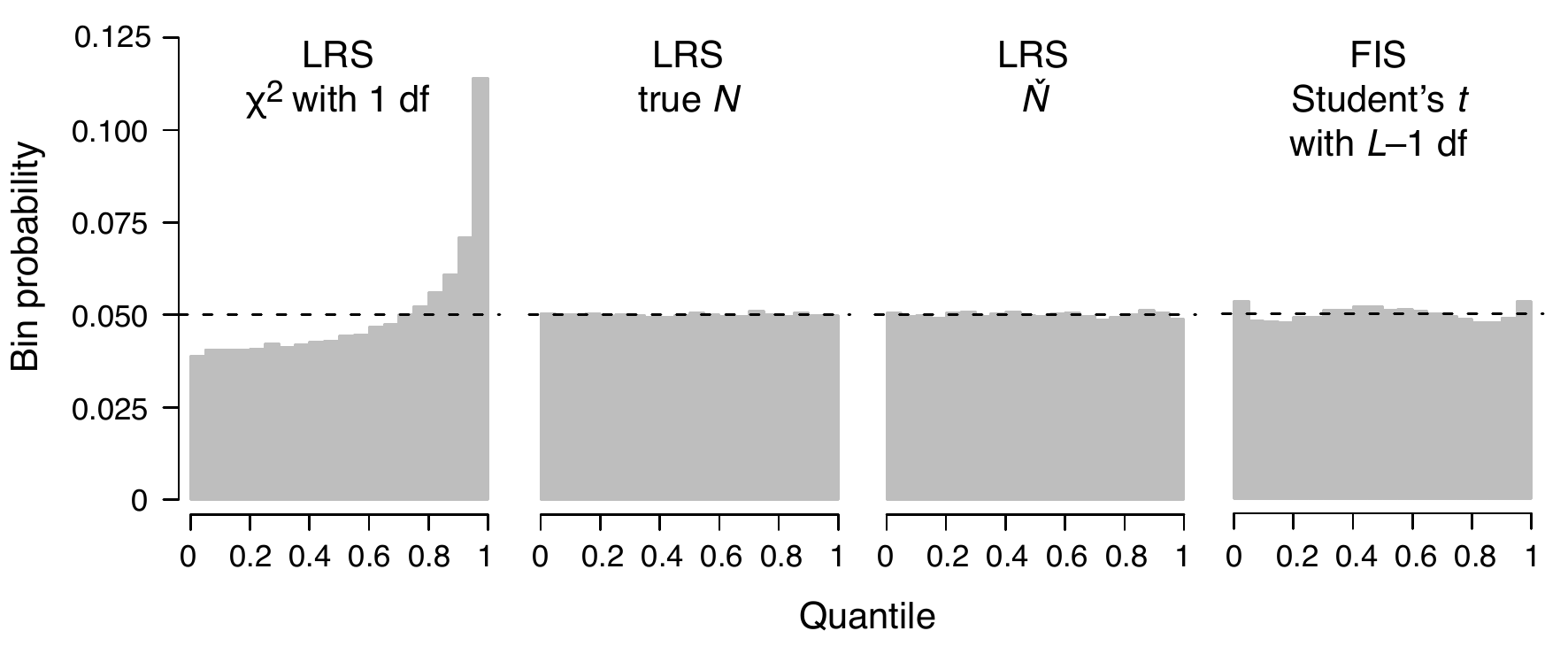}
\end{center}
\caption{
  {\bf Distributions of test statistics under the neutral null
  hypothesis.}
Histograms show the probabilities that the value of a test statistic generated
under the neutral null hypothesis falls within each vigintile (quantiles
of size 0.05) of another, approximate, distribution.  If the approximate
distribution is close to the true distribution the probability for each bin will
approximately equal 0.05 (dashed line).  The left three panels show the probability distributions
for the likelihood ratio statistic (LRS) to fall into the vigintiles of the $\chi^2$ distribution with 1 degree
of freedom, the
LRS distribution under the true $N$, and the empirical LRS distribution under $\check
N $, respectively.  The LRS falls in the top vigintiles of the
$\chi^2$ distribution more often than expected, indicating that the
$P$-value given by the $\chi^2$ distribution underestimates the probability of a
Type I error.  The distribution of LRS under the true $N$ is shown as a control
case.  The distribution of LRS under $\check N$ closely approximates the true LRS
distribution.  The rightmost panel shows the probabilities for the frequency
increment statistic (FIS) to fall
into each vigintile of the Student's $t$-distribution with $L-1$ degrees of freedom.  Student's $t$
is a good approximation for the true distribution of the FIS.  Parameter
values:  $N = 10^3$, $T = 100$, $\Delta=20$,  $L=5$, $\nu_0 = 0.5$; the number of
Wright-Fisher simulations was $3.5 \times 10^5$.
}
\label{fig:quantiles}
\end{figure}

  \begin{center}
    [Figure~\ref{fig:quantiles} approximately here]
    [Table 1 approximately here]
  \end{center}

%[AF:EDIT] from [Table 3 approximately here] to [Table 1 approximately here]

  The results of these analyses are shown in Figures~\ref{fig:quantiles} and S1, and in Tables~1 and S1.
  Figures 1 and S1 demonstrate that the $\chi^2$ distribution is a poor approximation for the true distribution of LRS under neutrality when the number of sampled time points is finite.
  If the LRS under the neutral null hypothesis followed the $\chi^2$ distribution,
  then the probability for LRS to fall into each vigintile of the $\chi^2$
  distribution would each equal 0.05.
  Instead, the LRS more often falls in the the top vigintiles of the $\chi^2$ distribution and, correspondingly, less often in the bottom vigintiles.
  This fact is problematic because it implies that the $P$-values calculated from
  the $\chi^2$ distribution will underestimate the probability of Type I error.
  Tables~1 and S1 show that this is indeed the case, even when as many as 100 time points are sampled.
  While the discrepancy between the actual probability of false positives and the
  $\chi^2$-based $P$-value is moderate (less than a factor of 2) for relatively
  high nominal $P$-values (e.g., above 1\%), the discrepancy becomes increasingly more severe for stringent $P$-values, 
  so that in some regimes the $\chi^2$ test rejects neutrality 50 times as often as it should (see Table~S1).

  The classical result of \citet{Wilks1938} guarantees that the LRS distribution
  will converge to the $\chi^2$ distribution as the number of data points increases.
  In our case, the LRS distribution should converge to the $\chi^2$ distribution with 1 df as the number of sampled points $L$ increases (and $\Delta$ decreases), while the time-series length $T$ remains constant.
  The $\chi^2$-based $P$-value should likewise converge to the true probability of Type I error.
  As expected, the values in columns 7 through 10 in the bottom section of Table~1
  and in the corresponding sections of Table~S1 approach 1 as $L$ increases.

  In addition to the deviation of the LRS distribution from the $\chi^2$
  distribution, the most likely population size under the null hypothesis, $\check
  N$, systematically overestimates the true population size, $N$, especially when
  the number of data points is small (see Tables~1 and S1). This phenomenon is consistent with previous reports \citep{Waples1989,WilliamsonSlatkin1999,Wang2001}.
  The bias in the inferred population size decreases with increasing number of
  data points, almost independently of the true population size or the observation time (see Figure S2).

\subsection{Two alternative tests of selection}

\subsubsection{The empirical likelihood ratio test (ELRT)}

  We propose two approaches to fix the shortcomings of the $\chi^2$ likelihood ratio test for
  selection in times series.
  The ideal approach would be to obtain the true distribution of the LRS
  by simulating the neutral Wright-Fisher model with the true population size, $N$.
  But since we are concerned with the case when $N$ is unknown,
  we propose to use the maximum-likelihood population size under neutrality,
  $\check N$, which we can estimate, in order to obtain the null distribution of the LRS.
  We call this approach the empirical likelihood ratio test (ELRT).

  Figure~\ref{fig:quantiles} shows that the LRS distribution generated under $\check N$ is an excellent approximation to the true LRS distribution, even when the number of sampled time points is small.  As a result, the $P$-values computed with the empirical LRS distribution provide an accurate description of the rate of false positive.
  Nevertheless, the ELRT approach suffers from two drawbacks, at least in its
  simplest implementation.  First, the Gaussian approximation that we
  employed to calculate the likelihoods becomes problematic in cases when the
  observed allele-frequency changes are large (for example if the allele is under
  very strong selection).  Large changes in allele frequency lead to small $\check N$ which leads to a high probability of absorption events, and the Gaussian approximation becomes inaccurate.
  Thus, somewhat paradoxically, we expect the ELRT based on the Gaussian approximation
  to lose power when the data come from populations under very strong selection.
  This problem is not intrinsic to the ELRT method, and indeed it could be
  remedied by calculating likelihoods using the (computationally intensive)
  diffusion approximation.  
  The second drawback of ELRT is that it is computationally intensive, even when
  using the fast Gaussian approximation for likelihoods. In particular,
  in order to obtain the approximate empirical LRS distribution, 
  many Wright-Fisher simulations must
  be performed, each accompanied by the calculation of the LRS.

  In the next section we propose another alternative to the $\chi^2$ likelihood ratio test that
  is computationally inexpensive, but somewhat less accurate than ELRT.

 %[AF:EDIT] intrinstic -> intrinsic 
 
\subsubsection{The frequency increment test (FIT)}
  We define the rescaled allele frequency increments as
  \begin{displaymath}
    Y_i = \frac{\nu_i - \nu_{i-1}}{\sqrt{2\nu_{i-1} (1-\nu_{i-1}) (t_{i} - t_{i-1})}}, \; i = 1,2,\dots, L.
  \end{displaymath}
  Since under the neutral null hypothesis the allele frequency $\nu$ behaves
away from the boundaries 0 or 1 approximately as Brownian motion (see
  \citet{Ewens} and Appendix), the random variables $Y_i$ are independent and approximately normally distributed with mean 0 and variance $1/N$ (see equations (\ref{eq:P_approx neut}), (\ref{eq:var(t) neut})).
  Under the alternative hypothesis, $Y_i$ are also independent and approximately
  normally distributed, but with a non-zero mean and a different variance (see equations~(\ref{eq:P_approx})--(\ref{eq:var(t)})).
  Thus, the problem of testing whether serial data come from a neutral
  population reduces to the problem of testing whether the rescaled allele
  frequency increments come from a normal distribution with mean zero (and unknown variance).
%
  %[AF:EDIT] right below
    %OLD:   The latter problem is one the most classical problems in statistics
    %NEW:   The latter problem is one of the most classical problems in statistics
%    
  The latter problem is one the most classical problems in statistics, and it has
  a well known and elegant solution: the $t$-test.
  The frequency increment statistic (FIS), defined as
  \begin{equation}
    t_\mathrm{FI}(\textrm{Data}) = \frac{\bar Y}{\sqrt{S^2/L}}, \label{eq:t-stat}
  \end{equation}
where $\bar Y$ and $S$ are the sample mean and the sample variance
  \begin{displaymath}
    \bar Y = \frac{1}{L}\sum_{i=1}^L Y_i \quad \textrm{and} \quad  S^2 = \frac{1}{L-1}\sum_{i=1}^L \left( Y_i - \bar Y \right)^2,
  \end{displaymath}
is distributed according to the Student's $t$-distribution with $L-1$ degrees of
freedom, under the neutral null hypothesis. Note that the unknown nuisance
parameter, $N$, in the population-genetic problem corresponds to the unknown
variance in the $t$-test.
  We call this test the frequency increment test (FIT).
  In addition to being simple and computationally trivial, this test is also {\em
  the most powerful similar test }\citep[see][Chapter 21]{StuartOrdArnold} of the
  selection hypothesis against the neutral null hypothesis, provided frequencies
  are far from the boundaries 0 and 1.

  Figure~\ref{fig:quantiles} and Table~2 show that FIT substantially outperforms
  the $\chi^2$ likelihood ratio test, in the sense that the nominal FIT $P$-value represents the probability of a Type I error more accurately 
  than does the $\chi^2$-based $P$-value.
%  
    %[AF:EDIT] right below
    %OLD:   Nevertheless, FIT $P$-value is not exact.
    %NEW:   Nevertheless, the FIT $P$-value is not exact.
 %   
  Nevertheless, the FIT $P$-value is not exact.
  Under most parameter regimes where the probability of Type I error deviates from
  the $P$-value reported by the $t$-distribution, FIT appears to be overly
  conservative (i.e., the $P$-value overestimates the probability of Type  I
  error), but how precisely this depends on $N$, $L$, $\Delta$ is complicated (Table~2).
  In any case, the inaccuracies in the probability of Type I error under the FIT
  are an order of magnitude smaller than those under $\chi^2$ LRT, in all parameter regimes tested.

  %[AF:EDIT] dependends -> depends 

  \begin{center}
    [Table~2 approximately here]
  \end{center}

\subsection{Power of ELRT and FIT to detect selection }

\begin{figure}[th]
\centering
\includegraphics{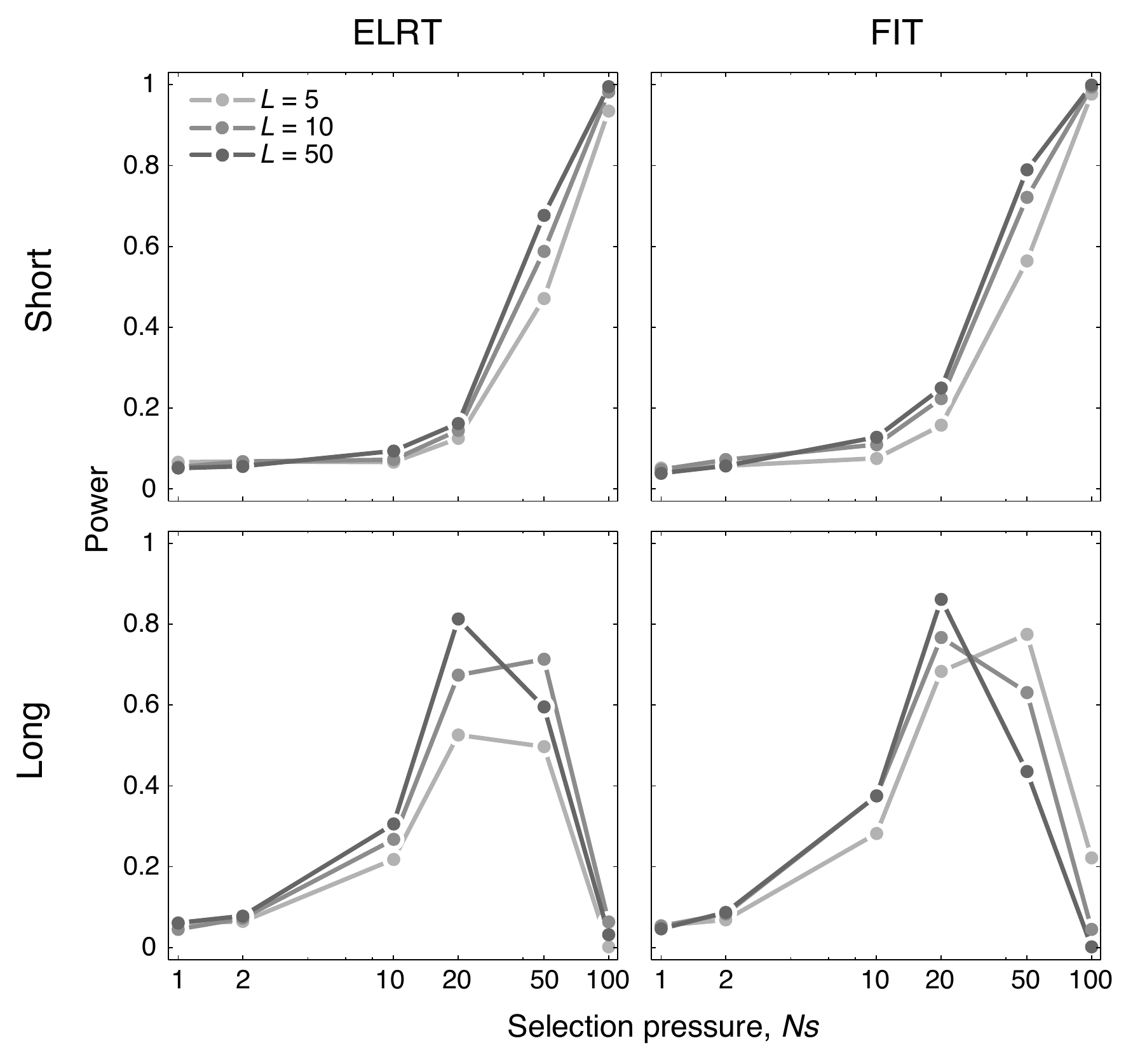}
\caption{
{\bf Power of ELRT and FIT to detect selection of different strength.}
  Power is reported as the fraction of trial data sets generated by the Wright-Fisher model with selection for which ELRT (left column) or FIT (right column) reject the neutral null hypothesis at $P$-value $\alpha=0.05$ 
  in short ($ T = 0.01 N $, top row) and ``long'' ($ T = 0.1 N $, bottom row) time series.
  Both tests gain power with increasing selection pressure, but in long time
  series they start to lose power when selection becomes very strong (see text for details).
  Power of both tests grows weakly with the number of sampled time points, $L$.
  We ran $10^3$ trials with $N=10^4$ and initial allele frequency $\nu_0 = 0.5$.
  Trials that produced absorption events within the sampling period were discarded.
}
\label{fig:power}
\end{figure}

  \begin{center}
    [Figure~\ref{fig:power} approximately here]
  \end{center}

  Next we determined the power of ELRT and FIT to detect selection in
  allele-frequency data, in terms of the strength of selection, time series length, and sampling frequency.
  To this end, we ran Wright-Fisher simulations  with population size $N = 10^4$
  as described above, but now with allele $A$ possessing selective advantage $s$.
  For each value of the scaled selection coefficient $Ns$ ranging from $1$ to $100$ we simulated $10^4$ allele-frequency trajectories and sampled from them in the first $T = N/100 = 100 $ or in the first  $T = N/10 = 1000 $ generations.
  These two sampling schemes gave rise to the ``short'' and ``long'' allele-frequency time series.
  For each time series, we sampled the frequencies of the selected allele at $L+1$ time points equally spaced $\Delta$ generations apart, with $L$ taking values 5, 10, or 50.
  For each combination of $s$, $T$, and $L$, we performed ELRT and FIT, and we calculated the frequency with which they reject neutrality at $P$-value 0.05.
  When computing the null distribution of LRS in the ELRT, we encountered some neutral Wright-Fisher trials that exhibited
  an absorption event during the observation period $T$.
  Instead of discarding such trials, we include them into the estimation of the empirical LRS distribution by conservatively assigning them to the maximum LRS value of the neutral trials.

\begin{figure}[th]
\centering
\includegraphics{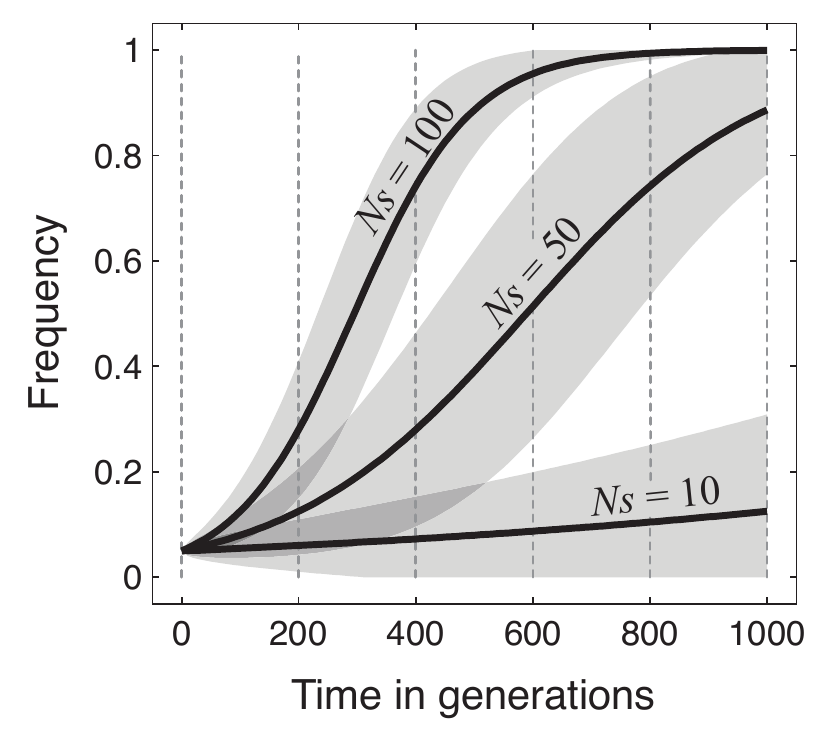}
\caption{
{\bf Schematic diagram describing the power of any test for selection in allele-frequency time-series data.}   
Thick black lines show the expected frequency dynamics (equation (\ref{eq:g(t)}))
of alleles with selection coefficients $s = 0.001, 0.005, 0.01 $, initiated at frequency $x_0 = 0.05$.
  Gray areas denote $\pm \sqrt{\sigma^2(t,x_0)/N} $, where $\sigma^2$ is given by equation (\ref{eq:var(t)}) and $N=10^4$, 
  which illustrate the size of stochastic fluctuations around the expected frequency.
  Vertical dashed gray lines show hypothetical sampling time points.
  When the selection coefficient is low ($Ns = 10$) stochastic fluctuations dominate, and tests of selection have low power.
  When selection coefficient is high ($Ns = 100$) fixation events occur within the sampling interval and some sampling points 
  (at 800 and 1000 generations) become uninformative, which also leads to loss of power.
  For a given sampling interval $T$ power is maximized for intermediate selection coefficients ($Ns = 50$).
}
\label{fig:sweeps}
\end{figure}

  Figure~\ref{fig:power} shows that both tests posses substantial power to detect moderate to strong selection ($Ns > 10$), but they lose power when
  selection is very strong.  As illustrated in Figure~\ref{fig:sweeps}, such behavior is expected for any test of selection from time series data.
  Consider a fixed sampling duration $T$.  Clearly, if selection is very weak, it will not be able to change the allele frequency substantially during
  this time interval, and so the observed allele-frequency changes will be dominated by noise.  On the other hand, when selection is very strong, the
  allele will go to fixation within the interval $T$, and so some of the samples in the later part of the interval will carry no information about the
  allele dynamics.  For example, in Figure~\ref{fig:sweeps}, an allele with selection coefficient $Ns = 100$ typically fixes in less 800 generations,
  and so samples taken after generation 800 are uninformative.  In the extreme case of very strong selection, the allele will fix between the first
  and second sampling time points. In this case, without knowledge of the population size, we could not determine whether the time series was caused
  by strong selection or strong genetic drift. Thus, any test of selection based on time series data will loose power for either very weak or very
  strong selection pressures.

  \begin{center}
    [Figure~\ref{fig:sweeps} approximately here]
  \end{center}

The intuition outlined above suggests that a given sampling interval $T$ sets the scale for selection coefficients that we have power to detect, $s_\mathrm{power}(T)$.
  We can estimate $s_\mathrm{power}(T)$ by inverting the logic of this intuition:
  for selection strength $s$, there is an optimal sampling interval that maximizes
  the power of tests to detect this selection.
  Such sampling interval should be long enough for selection to substantially change the allele frequency but short enough to avoid fixation.
  From equation (\ref{eq:g(t)}), the expected time $t(x_f,x_0; s)$ it takes for an allele with selection coefficient $s$ to reach frequency $x_f$ from the initial frequency $x_0$ is approximately given by
  \begin{displaymath}
    t(x_f,x_0; s) = \frac{1}{s} \ln\left(
      \frac{x_f}{1-x_f} \frac{1-x_0}{x_0} 
    \right).
  \end{displaymath}
  Setting $t(x_f, x_0; s_\mathrm{power}) = T $ with some arbitrary $x_f$ close to
  1, we predict that tests of selection in a time series of length $T$ will have the maximal power to detect selection coefficients on the order of
  \begin{displaymath}
    s_\mathrm{power}(T) = \frac{1}{T} \ln\left(
      \frac{x_f}{1-x_f} \frac{1-x_0}{x_0}
    \right).
  \end{displaymath}
  Setting $x_0 =0.5$ as in our simulations and $ x_f = 0.95 $ (this choice is
  arbitrary and not critical for determining the order of magnitude of
  $s_\mathrm{power}$), we predict that tests of selection will have maximal power
  to detect selection of strength $ s_\mathrm{power} = 0.029 $ in time series of
  length $T=100$ generations; and $s_\mathrm{power} = 0.0029 $ in time series of length $ T = 1000$ generations.
  For a population of size $N = 10^4$, this translates into $Ns_\mathrm{power} = 290$ and $Ns_\mathrm{power} = 29 $, respectively, 
  which is consistent with our numerical results (Figure~\ref{fig:power}). These
  power calculations are generic properties of any test of selection in time-series data.

  ELRT and FIT have an additional complication in that they cannot be applied to data points after an absorption event.
 % SK: removed "(which are uninformative in any case)"
%
  In plotting Figure \ref{fig:power}, we discarded all trials in which an absorption event occurred within the sampling period, even though some of these trials likely had a detectable signature of selection prior to the absorption event.
  Thus, Figure~\ref{fig:power} shows the lower bound on the power of our tests.

Aside from these gross properties of power, we found that FIT has slightly more power than ELRT, and that power of both tests increases weakly with the number of sampled time points $L$, with all other parameters being equal.

\subsection{The effects of noisy sampling}

\begin{figure}[th]
\centering
\includegraphics{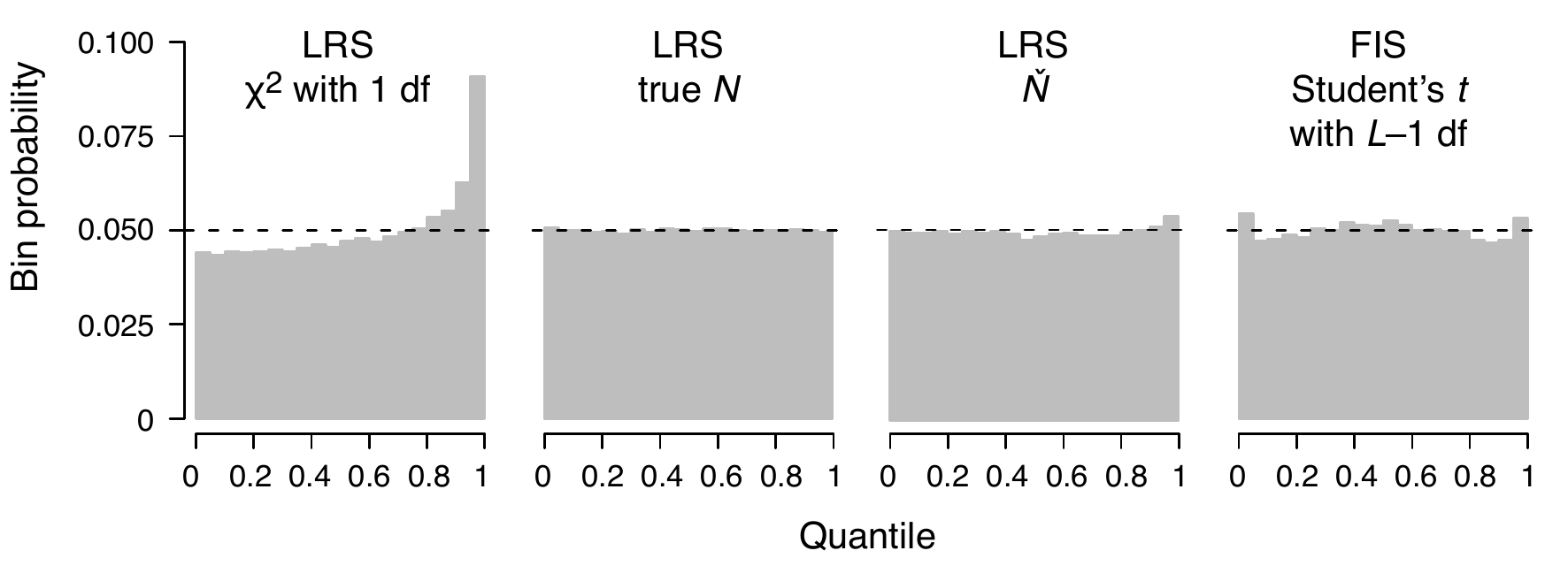}
\caption{
  {\bf Distributions of test statistics under the neutral null hypothesis,
  when allele frequencies are sampled with noise.}
  Histograms show the probabilities that the value of a test statistic generated under the neutral null hypothesis falls within each of the 
  vigintiles (quantiles of size 0.05) of another, approximate, distribution.
  Notations are as in Figure~\ref{fig:quantiles}.
  Parameter values:  $N = 10^3$, $T = 100$, $\Delta=20$,  $L=5$, $\nu_0 = 0.5$, $n=500$; the number of Wright-Fisher simulations was $2\times 10^5$.
}
\label{fig:quantiles sampled}
\end{figure}

  \begin{center}
    [Figure~\ref{fig:quantiles sampled} approximately here]
  \end{center}

So far we have studied tests of selection assuming that allele frequencies are measured with (perfect) accuracy in successive time points. In this
section, we investigate the behavior of the FIT and ELRT in a more realistic situation -- when allele frequencies are estimated, at each time
point, by sampling a limited number of individuals from the population and typing them with respect to the focal locus. To study this, we used the
same simulated time-series trajectories as in previous sections, but instead of analyzing the true allele frequencies, $x$, we drew binomial random
variables with sample size $n$ and success probability $x$ to obtain the sampled allele frequencies $\nu$.  We then analyzed the test size and power
treating the  sampled allele frequencies $\nu$ as the data.
  
  %[AF:EDIT] below: overeestimate -> overestimate 
  
\begin{figure}[th]
\centering
\includegraphics{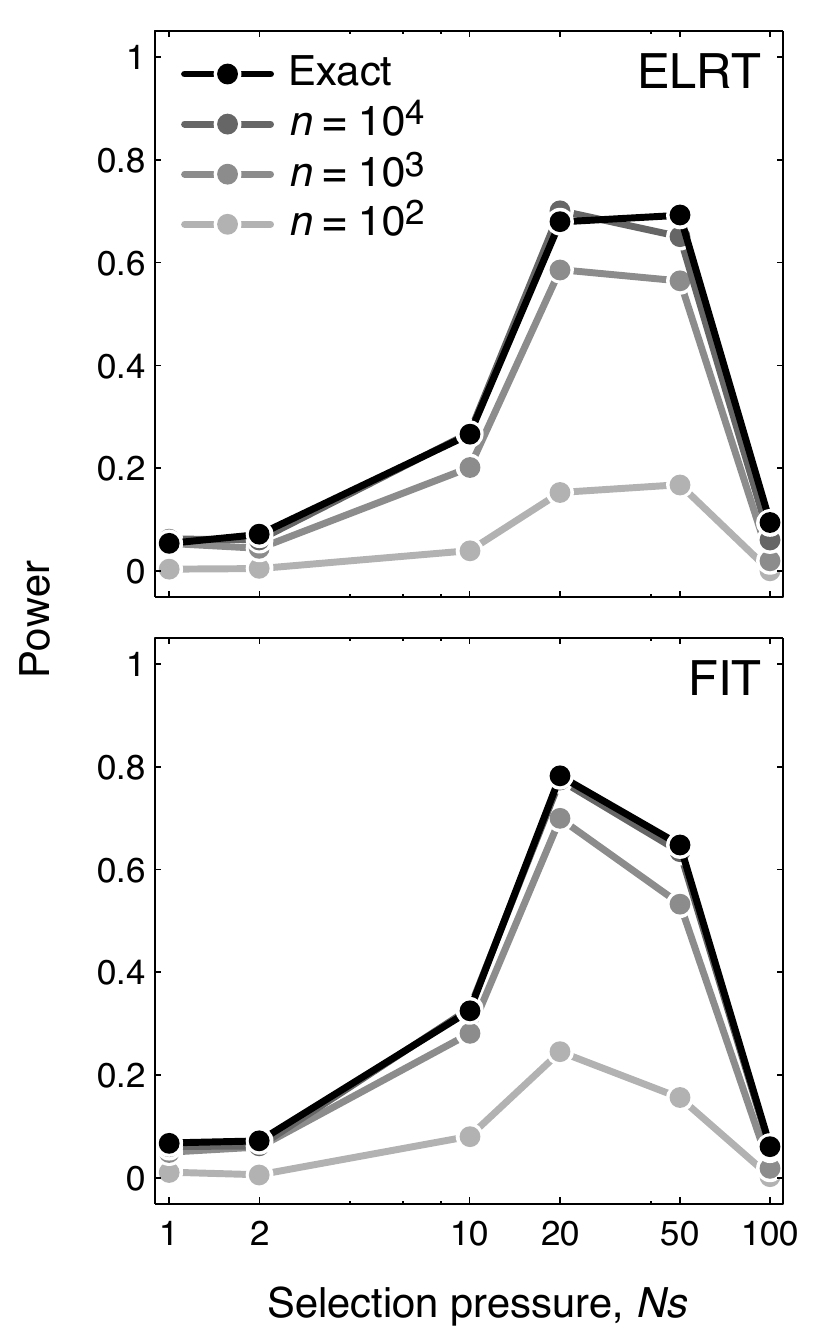}
\caption{
{\bf Power of ELRT and FIT to detect selection of different strengths, under various sampling regimes.}
  Parameter values:  $N = 10^4$, $T = 1000$, $\Delta=100$,  $L=10$, $\nu_0 = 0.5$; the number of Wright-Fisher simulations was $10^3$.
}
\label{fig:power sampled}
\end{figure}

As shown in Figure~\ref{fig:quantiles sampled},  when sample sizes are sufficiently large ($n = 500$) the $P$-values produced by ELRT and FIT remain
accurate representations of the true Type I error probability.  When the sample sizes become too small ($n \leq 100$), both tests become overly
conservative, i.e., the $P$-value produced by ELRT and FIT overestimate the probability of Type I error (see Figure~S3).  Note that the LRT 
also becomes overly conservative in this regime, even if the $\chi^2$ distribution or the distribution of LRS under true $N$ are used (Figure~S3).  This in
itself is not problematic and it simply implies that the $P$-values from such tests should be viewed as upper bounds on the actual probability of Type I error.  More problematic is the associated decline in power of both tests as samples size $n$ decreases (Figure~\ref{fig:power sampled}).
%   When the number of samples falls below 100 per sampled time point, i.e., when sampling noise overwhelms demographic noise, we lose the ability to
%   detect selection unless the selection coefficient is very high (e.g., $Ns>20$).
The dependence of power on the strength of selection in the presence of sampling noise remains the same as in the absence of sampling noise, with the
power curves shifted downwards (Figure~\ref{fig:power sampled}).

  \begin{center}
    [Figure~\ref{fig:power sampled} approximately here]
  \end{center}

\subsection{Applications to empirical data}

  In this section, we apply our tests of selection to
  allele-frequency time series from three previously published experimental data
  sets, as well as some additional new experimental data.

\subsubsection{Bacteriophage evolved at high temperature}
  The first data set is from an experiment described by \citet{BollbackHuelsenbeck2007}.
  \citet{BollbackHuelsenbeck2007} evolved three lines of bacteriophage MS2, which
  infects {\em Escherichia coli, }at increasingly high temperatures, from
  39\textdegree C to 43\textdegree C.
  After 50 passages, each corresponding to approximately three bursts, they identified mutations that were 
  segregating in the populations and determined the frequencies of these mutations at the previous time points. 
  From this data set we selected allele-frequency trajectories that remained at
  intermediate frequencies between 0 and 1 for at least two consecutive time
  points and applied FIT, but not ELRT (Table~3).
  We could not apply ELRT to these data for two reasons.
  First, some time series had only two time points at which the mutant allele was at intermediate frequencies.
  The maximum-likelihood approaches cannot estimate both $N$ and $s$ in such cases (see Materials and Methods).
  Second, the frequencies of the remaining alleles changed so fast (e.g., from 30\% to 90\% in 10 passages) that the ML-estimated population sizes under neutrality, $\check N$, were very small (see Table~3), and so neutral simulations were dominated by absorption events.
  % Second, for alleles observed at intermediate frequencies for three or more time points, the ML-estimated population sizes under neutrality were very small (see Table~3), and so neutral simulations were dominated by absorption events.
  
    %[AF:EDIT] above: segragating -> segregating 
    %[AF:EDIT] above: absorbtion -> absorption 
    
  When we applied FIT to these data, we found that only one time series
  produced a significant $P$-value (mutation C3224U in line 3), despite the fact that most of the identified mutations are likely to be beneficial.
  The poor performance of our tests on these data is expected for two reasons.
  First, the sample sizes in these data set are very small ($ n \leq 10$), and we expect our tests to have very low power.
  Second, even though all mutations are probably beneficial, not all frequency
  trajectories are monotonically increasing, and some of them are even decreasing (e.g., mutation C1549U/A in line 3), 
  presumably due to clonal interference \citep{GerrishLenski1998}, which further
  reduces the power of our test.

  \begin{center}
    [Table 3 approximately here]
  \end{center}

\subsubsection{Deep population sequencing of adapting yeast populations}

\begin{figure}[th]
\centering
\includegraphics{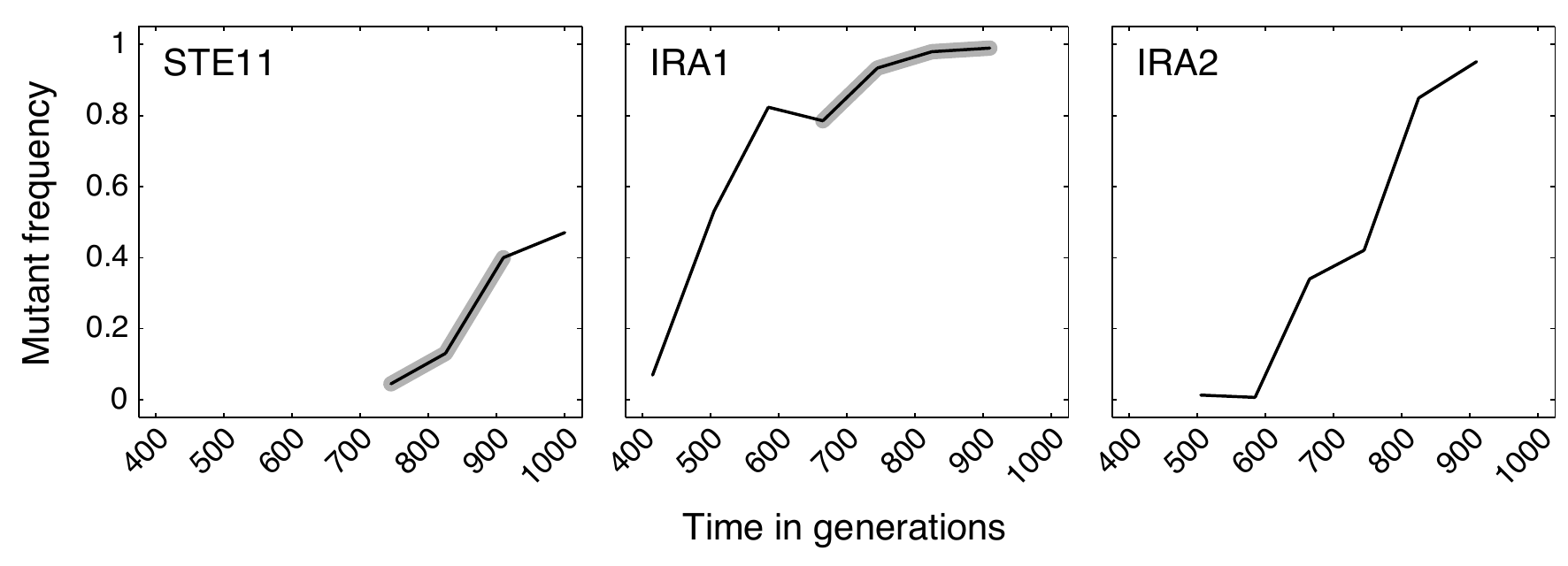}
\caption{ {\bf Application of ELRT and FIT to allele-frequency time series from Lang et
al.}
Each panel shows the estimated frequency of a mutant allele in the long-term
evolution lines described in~\citet{LangBotsteinDesai2011,LangRice2013}: left
panel shows the frequency of mutation D579Y in gene {\it STE11 }in population RMB2-F01;
middle panel shows the frequency of mutation Y822* in gene {\it IRA1 }in population
RMS1-D12; right panel shows the frequency of mutation A2698T in gene {\it IRA2 }in
population BYS2-D06. Gray shading highlights the data points for which FIT and ELRT
identify selection.
}
\label{fig:data}
\end{figure}

The second data set we analyzed is from an experiment in which \citet{LangBotsteinDesai2011} evolved 592 populations of yeast {\em Saccharomyces
cerevisiae }in  rich medium for 1000 generations.  The original experiment tracked the appearance and fate of sterile mutations which are known to be
beneficial under the chosen experimental conditions \citep{LangBotsteinDesai2011}.  Subsequently, some of these populations were deep-sequenced, and
many other adaptive mutations were identified \citep{LangRice2013}.  From this large data set, we selected three allele frequency trajectories of
mutations in genes {\em STE11, IRA1, }and {\em IRA2 }which arose in three different populations (Figure~\ref{fig:data}, Table~S2).  Applying ELRT and
FIT to these time series, we found that our tests return best results when used on subsets of each time series (Figure~\ref{fig:data}, Table~S2).
Based on these truncated time series, both ELRT and FIT identified that the trajectories of the mutant {\em STE11 }and {\em IRA1 }alleles, but not
that of the mutant {\em IRA2 }allele, were positively selected.  Given the knowledge that the experimental population sizes exceed $10^4$ individuals,
and the fact that mutations in genes {\em STE11, IRA1, }and {\em IRA2 }independently arose and spread in several parallel lines, it is likely that all
three mutations are in fact beneficial \citep{LangRice2013}.  Our tests do not take these two critical pieces of information into account, but they are
still able to identify the action of positive selection in two out of three cases, based solely on allele frequencies estimated 
from samples of size $n \leq
150$.
  
      %[AF:EDIT] above
    %OLD:  Our tests do not take these two critical pieces information into account, but they are
    %NEW:  Our tests do not take these two critical pieces of information into account, but they are
    % independengly -> independently 
  
\subsubsection{Yeast populations evolved at different population sizes}

The third data set we analyzed is from an experiment performed by one of us (SK) and described in Ref. \citep{Kryazhimskiy2012}.  In this experiment,
1008 populations of yeast {\em Saccharomyces cerevisiae } were evolved under conditions similar to those in the experiment by
\citet{LangBotsteinDesai2011}, but under various population sizes and migration regimes.  After 500 generations of evolution, fitnesses of these
populations were measured in a competition experiment.  Fitness data for 976 of these populations were previously described in
\cite{Kryazhimskiy2012}.  Here, we analyzed the published competition assays from 736 well-mixed populations, referred to as ``No'', ``Small WM'', and
``Large WM'', as well as unpublished data from additional 32 well-mixed populations of intermediate size referred to as ``Medium WM''.  All these
populations were evolved in exactly identical conditions, except for the serial transfer bottleneck size.  In particular, the bottleneck size was
approximately $10^3$ individuals in ``No'' populations, and 5, 10, and 20 times larger than that in ``Small WM'', ``Medium WM'', and ``Large WM'',
respectively.  The fitnesses of all populations were measured in competition assays with at least three-fold replication.
  As described in Ref. \citep{Kryazhimskiy2012}, each competition assay consists of measuring the frequency of the evolved population relative to a fluorescently labelled reference strain at two time points.
  The raw flow cytometry counts for all populations used here (including those published previously) are reported in Table~S4.

  As mentioned in Materials and Methods, when the time series contains only two time points, there is not enough information to estimate the population size (or, equivalently, the variance of the distribution of frequency increments).
  However, FIT can be easily applied to frequency increments pooled across replicate fitness measurements.
  In particular, if $\nu_{ki}$ is the frequency of the evolved population at time point $i$ (with $t_0 =0 $ and $t_1 = 20 $) in replicate assay $k$ (with $k=1,\dots, K$), then we define the frequency increment in replicate $k$ as
  \begin{displaymath}
    Y_k = \frac{\nu_{k1} - \nu_{k0}}{\sqrt{2\nu_{k0} (1-\nu_{k0}) (t_1 - t_0)}},
  \end{displaymath}
and calculate the frequency increment statistic according to equation (\ref{eq:t-stat}), with $L$ replaced by the number of replicates $K$.

  The results of FIT applied to these data are reported in Table S3 and summarized in Table~4.
  We find that FIT rejects the neutral null hypothesis at various stringency cutoffs for all ``Medium WM'' and ``Large WM'' populations and for the majority of ``Small WM'' populations.
  At the same time, FIT rejects neutrality for only $\sim 34$\% of ``No'' populations at the $P$-value cutoff 0.05, and only $\sim 1$\% of  ``No'' populations at the $P$-value cutoff of 0.001.
  In both of these cases the observed numbers of positives significantly exceeds the numbers of false positives expected due to multiple testing.
  These results demonstrate that FIT reliably detects the action of natural selection in data from microbial evolution experiments.
  Moreover, since we do not know which populations truly adapted in this experiment, these results  inform us that, when the bottleneck size exceeding 5000 individuals, nearly all populations undergo significant adaption during 500 generations of evolution, but when the bottleneck size is 1000, only about $34$\% of populations do so.
  These results are consistent with the expectation that larger populations adapt faster and suffer less from the accumulation of deleterious mutations, compared to small populations.

  \begin{center}
    [Table 4 approximately here]
  \end{center}

\section{Conclusions}
  We have shown that the standard $\chi^2$-based test for selection in time series of allele frequencies \citep{Bollback2008} is subject to a 
  greatly elevated false discovery rate in the practical regime of relatively few sampled time points.
  As a result of this bias, the $\chi^2$ LRT is not a reliable test for selection
  in many practical time series, 
  because its $P$-value underestimates the rate of false positives, especially when the allele frequencies are measured accurately.
  We proposed two new tests to address this problem, and we showed that both of them accurately 
  estimate the probability of Type I error and have power to detect selection in parameter regimes that are reasonable for 
  many evolution experiments and natural populations.

  Our tests were initially developed under the assumption that sampling noise is
  negligible and that the estimated allele frequencies can be treated as exact.
  In many situations, such as microbial laboratory experiments, this assumption is not restrictive.
  Indeed, when allele frequencies are measured with high-throughput methods such as flow cytometry \citep{LangBotsteinDesai2011,Kryazhimskiy2012} or deep population sequencing \citep{Smith2011,LangRice2013}, the sample sizes often exceeds the population size.  
  On the other hand, when samples are derived from natural populations, this assumption is likely to be violated.
  In this case, our tests remain conservative, but lose power to detect selection, especially when selection is weak.
  This is expected because when sampling noise dominates demographic stochasticity the information about the population size that is contained in allele-frequency fluctuations is lost.
  In principle, the population size can be inferred even in the presence of high sampling noise, if the time series is long enough.
  Indeed, if large frequency fluctuations are caused by low population size, time to absorption will be short, but if they are caused by sampling noise, time to absorption will be long.
  Moreover, incorporating time to absorption into tests of selection in time-series data would alleviate the ascertainment bias that arises when, for example, only those alleles are analyzed that reach sufficiently high frequencies in the population.
  % SK. OLD: Incorporating this information into tests of selection in time-series data is a subject of future research.

  The methods proposed here, just as the earlier $\chi^2$-based test, are limited to the regime in which the frequencies of alleles observed at a locus are not influenced by mutations that may arise elsewhere in the genome during the time of observation.
  Thus, our tests are perhaps most readily applicable to selection scans in full-genome time-series data like those now actively generated in evolution experiments in {\em Drosophila }\citep{Burke2010,Orozco2012}.
  It will also be applicable for asexual organisms when clonal interference is absent or weak, for example in competitive fitness assays \citep{Lenski1991,Gallet2012,Kryazhimskiy2012} or in tracking known polymorphisms in natural populations for a relatively short time \citep{Barrett2008,Winters2012,Pennings2013}.
  By contrast, inferring selection coefficients when allele dynamics are
  influenced by multiple linked sites is a substantially more difficult problem,
  which has begun to be addressed elsewhere
  \citep{IllingworthMustonen2011,Illingworth2012}, although not within
  the same rigorous population-genetic framework that treats all genotypic
  dynamics stochastically.

\section{Acknowledgements}

  The authors thank Todd Parsons for help with the Gaussian approximation, Daniel P. Rice for providing experimental data, and two anonymous reviewers for thoughtful comments.
  SK was supported by the Burroughs Wellcome Fund Career Award at Scientific Interface.

\section*{Appendix. Gaussian approximation to the Moran process}

  We approximate the continuous-time Moran processes with a combination of a deterministic process and Gaussian noise process.
  We follow here the procedure outlined by \citet{Pollett1990}, which is based on the results by \citet{Kurtz1970,Kurtz1971}.
  The Gaussian approximation used here is slightly different from that described by \citet{Nagylaki1990} in that (a) it does not assume that selection is weak and (b) allows for the values of the original and limiting processes at the initial time point to be different.
  %SK: added ref to Kurtz and Nagylaki.

  The Moran's stochastic process describes the number $n^{(N)}(t)$ of mutants in a population of constant size $N$ at time $t$.
  This number can increase by one from $i$ to $i+1$ with rate
  \begin{displaymath}
    r^{(N)}(i, i+1) = \mu_m i \frac{\lambda_w (N-i)}{\lambda_w (N-i) + \lambda_m i}
  \end{displaymath}
and decrease by one with rate
  \begin{displaymath}
    r^{(N)}(i, i-1) = \mu_w (N - i) \frac{\lambda_m i}{\lambda_w (N-i) + \lambda_m i}.
  \end{displaymath}
  Here,
  $ \mu_w $ and $ \lambda_w $ are the birth and death rates of the wildtype, and
  $ \mu_m $ and $ \lambda_m $ are the birth and death rates of the mutant type, respectively.
  We assume $ \lambda_w = \lambda_m $, $\mu_w = 1$, and let $ \mu_m = (1 + s) \mu_w = 1+s $.
  Then
  \begin{equation}
    r^{(N)}(i, i+1) = N f_{+1}(i/N), \quad 
    r^{(N)}(i, i-1) = N f_{-1}(i/N) 
    \label{eq:density-dep}
  \end{equation}
with
  \begin{displaymath}
    f_{+1}(x) = (1 + s) x (1 - x), \quad
    f_{-1}(x) = x (1 - x).
  \end{displaymath}
Define
  \begin{eqnarray*}
    F(x) & = & \sum_{\delta \in \{-1,+1\}} \delta f_{\delta}(x) = s x(1-x) \\
    G(x) & = & \sum_{\delta \in \{-1,+1\}} \delta^2 f_\delta(x) = (2+s)x(1-x).
  \end{eqnarray*}
  Let $X^{(N)}(t) = n^{(N)}(t) / N $ be the frequency of the mutant in the population at time $t$.
  The limit of $X^{(N)}$,
  $ g(t, x_0) = \lim_{N \to \infty} X^{(N)}(t) $,
is a deterministic function that, under certain regularity conditions, satisfies equations (\ref{eq: dg/dt}),~(\ref{eq: g(0)}) with $x_0 = \lim_{N \to \infty} X^{(N)}(0)$ and solution given by (\ref{eq:g(t)}).

  Now let
  \begin{equation}
    \label{eq:def:Z}
    Z(t) = \lim_{N\to\infty} \sqrt{N} \left( X^{(N)}(t) - g(t, x_0) \right)
  \end{equation}
be the asymptotic process that describes the noise around the deterministic trajectory.
  If we knew the distribution of $Z(t)$, we could approximate the frequency $X^{(N) }$ at a finite $N$ by
  \begin{equation}
    X^{(N)}(t) \approx g(t, x_0) + \frac{1}{\sqrt{N}} Z(t).
    \label{eq:approx0}
  \end{equation}

  The asymptotic noise process is in general a diffusion process, but, as long as it remains far from absorbing boundaries, it can be approximated by a Gaussian process with the corresponding first two moments.
  The advantage of this approach is that the first two moments of the diffusion process can be computed analytically, resulting in an expression for the probability distribution of the allele frequency at time $t$.

  If $ z_0 = \lim_{N \to \infty }\sqrt{N}(X^{(N)}(0) - x_0) $ is the initial value of the limiting noise process, then the mean and variance of the noise process at time $t \geq 0 $ are $  {\mathbb E}Z(t) = M(t, x_0) z_0 $ and $\mathrm{Var} \; Z(t) = \sigma^2(t, x_0)$ respectively, where $M(t,x_0)$ satisfies the equations
  \begin{eqnarray}
    \frac{d M}{dt} & = & F'(g(t, x_0)) M = s \frac{(1-x_0)e^{-st} - x_0}{(1-x_0)e^{-st} + x_0} M \label{eq: dM/dt} \\
    M(0, x_0) & = & 1 \label{eq: M(0)}
  \end{eqnarray}
and $\sigma^2(t, x_0) $ satisfies the equations
  \begin{eqnarray}
    \frac{d\sigma^2}{dt} & = & 2 F'(g(t,x_0)) \sigma^2 + G(g(t,x_0)) = \nonumber \\
      & = & \frac{(1-x_0)e^{-st} - x_0}{(1-x_0)e^{-st} + x_0} \; 2 s \sigma^2
      + \frac{(2+s)x_0(1-x_0)e^{-st}}{((1-x_0)e^{-st} + x_0)^{-2}}  \label{eq: dsigma/dt} \\
    \sigma^2(0, x_0) & = & 0. \label{eq: sigma(0)}
  \end{eqnarray}
  The solution to equations (\ref{eq: dM/dt}),~(\ref{eq: M(0)}) is given by
\begin{displaymath}
  M(t, x_0) = \exp\left\{ \int_0^t F'(g(\tau, x_0)) d\tau \right\},
\end{displaymath}
that, after substituting $F'$ and $g $, yields
\begin{displaymath}
      M(t, x_0) = e^{-s t} \left(x_0 + (1 - x_0) e^{-s t} \right)^{-2}.
\end{displaymath}
  The solution to equations (\ref{eq: dsigma/dt}),~(\ref{eq: sigma(0)}) is given by
\begin{displaymath}
  \sigma^2(t, x_0) = M^2(t, x_0) \int_{0}^t M^{-2}(\tau, x_0) G(g(\tau, x_0)) d\tau,
\end{displaymath}
that, after substituting $ G $ and $ g $, yields
\begin{eqnarray*}
      \sigma^2(t, x_0) & = & M^2(t, x_0) (2 + s) x_0 (1-x_0)s^{-1}  \nonumber \\
    & \times & \Big[
      2x_0(1-x_0) s t + x_0^2e^{s  t}-(1-x_0)^2e^{-s  t} + (1-x_0)^2 -x_0^2
    \Big].
\end{eqnarray*}
  If the true state of the stochastic process $X^{(N)}$ is known to be $X^{(N)}(0)$ at the time point 0, we can approximate the initial value of the limiting noise process as $z_0 \approx \sqrt{N}(X^{(N)}(0) - x_0)$.
  Then from (\ref{eq:approx0}) we have
  \begin{eqnarray*}
    {\mathbb E} X^{(N)}(t) & \approx & 
    g(t, x_0) + M(t, x_0) \left( X^{(N)}(0) - x_0 \right),  \\
    \mathrm{Var} \; X^{(N)}(t) & \approx & \frac{1}{N}\sigma^2(t, x_0).
  \end{eqnarray*}
  Analogously, if the value of the process $X^{(N)}$ is known to be $X^{(N)}(t^\prime) $ at a later time $t^\prime \geq x_0$, then at time $t \geq t^\prime$ we have
  \begin{eqnarray}
    {\mathbb E}_{t^\prime} X^{(N)}(t) & \approx & 
    g(t, x_0) + M(\Delta t, g(t^\prime, x_0)) \left( X^{(N)}(t^\prime) - g(t^\prime, x_0) \right), \label{eq:approx mean} \\
    \mathrm{Var}_{t^\prime} \; X^{(N)}(t) & \approx & 
    \frac{1}{N}\sigma^2(\Delta t, g(t^\prime, x_0)),
    \label{eq:approx var}
  \end{eqnarray}
where $\Delta t = t-t^\prime$, and ${\mathbb E}_{t^\prime}$ and $\mathrm{Var}_{t^\prime}$ denote conditional expectation and variance given the state of the process at time $t^\prime$.
  Thus, the conditional distribution of the allele frequency $ X^{(N)} $ at time $t$ given its value at time $t^\prime \leq t$ can be approximated by a Gaussian distribution with mean given by (\ref{eq:approx mean}) and variance given by (\ref{eq:approx var}).
  We apply this approximation to every observation interval $(t_{i-1}, t_{i}), i = 1, \dots, L$.
  As noted above, the initial value of the deterministic process, $x_0$, is a free parameter that can be fitted along with $N$ and $s$.
  However, we set $x_0 $ to be equal to the observed allele frequency $\nu_0$ at time 0 in order to reduce the number of fitted parameters.

  Note that the approximations described here work for the Moran process which is density-dependent as can be seen from equations (\ref{eq:density-dep}).
  The Wright-Fisher process is not density-dependent and, strictly speaking, the approximations described here are not valid, although in practice they work well.

%\bibliographystyle{genetics}
%\bibliography{ref_exp_evol,ref_exp_evol_dyn,ref_time-series,ref_various}

\clearpage

\section*{Figures and tables}

\clearpage

%[AF:EDIT] added 'the' to 'the number of Wright-Fisher simulations...'
%[AF:EDIT] Fixed spelling of frequency

%\clearpage

%\clearpage

%\clearpage
%[AF:EDIT] added 'the' to 'the number of Wright-Fisher simulations...'
%[AF:EDIT] varous -> various 

%[AF:EDIT] $n=500$ removed from figure caption. 
%[AF:EDIT] added 'the' to 'the number of Wright-Fisher simulations...'

%\clearpage

\clearpage

\begin{table}[th]
\centering
\includegraphics{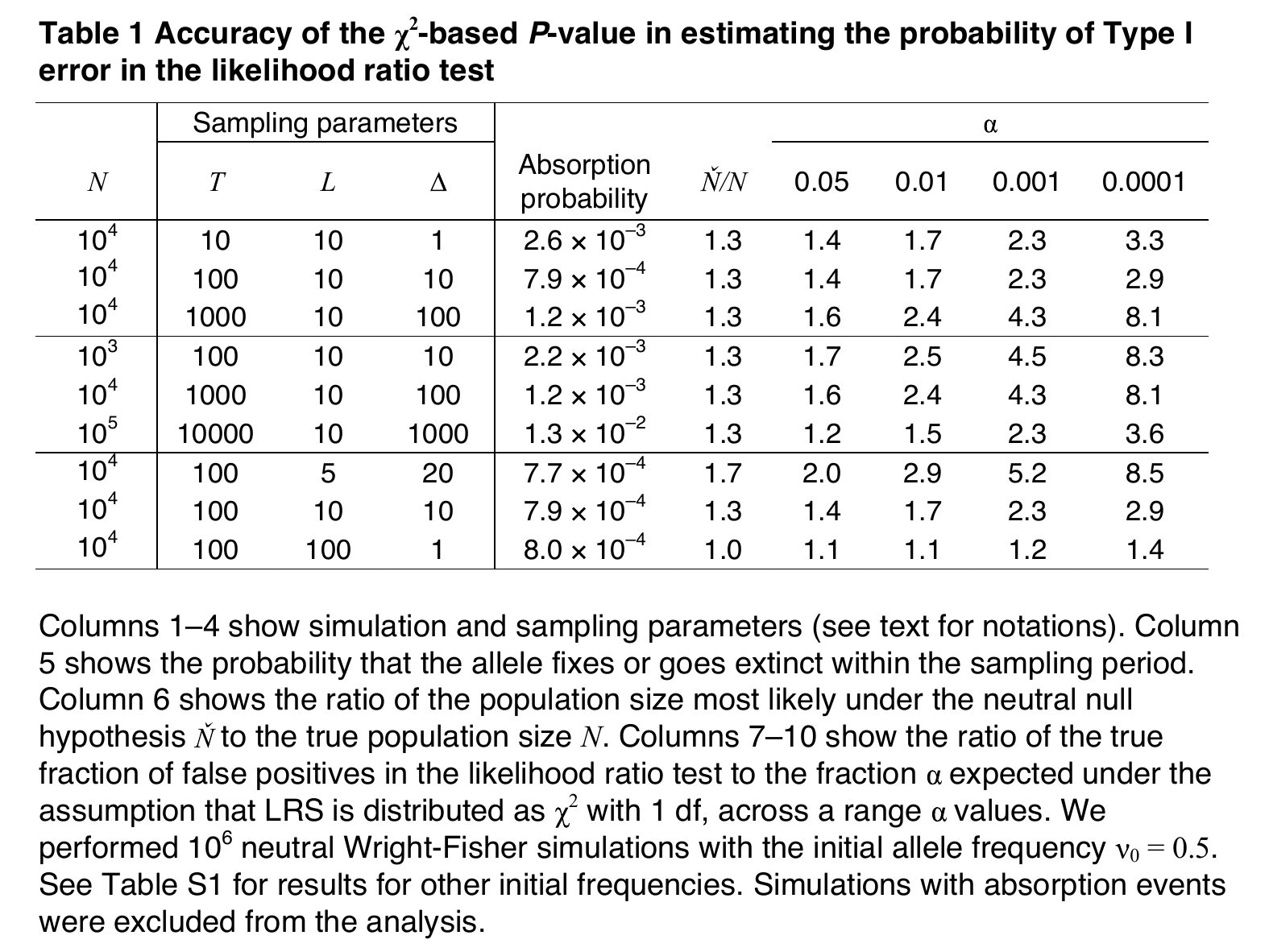}
\end{table}

\begin{table}[th]
\centering
\includegraphics{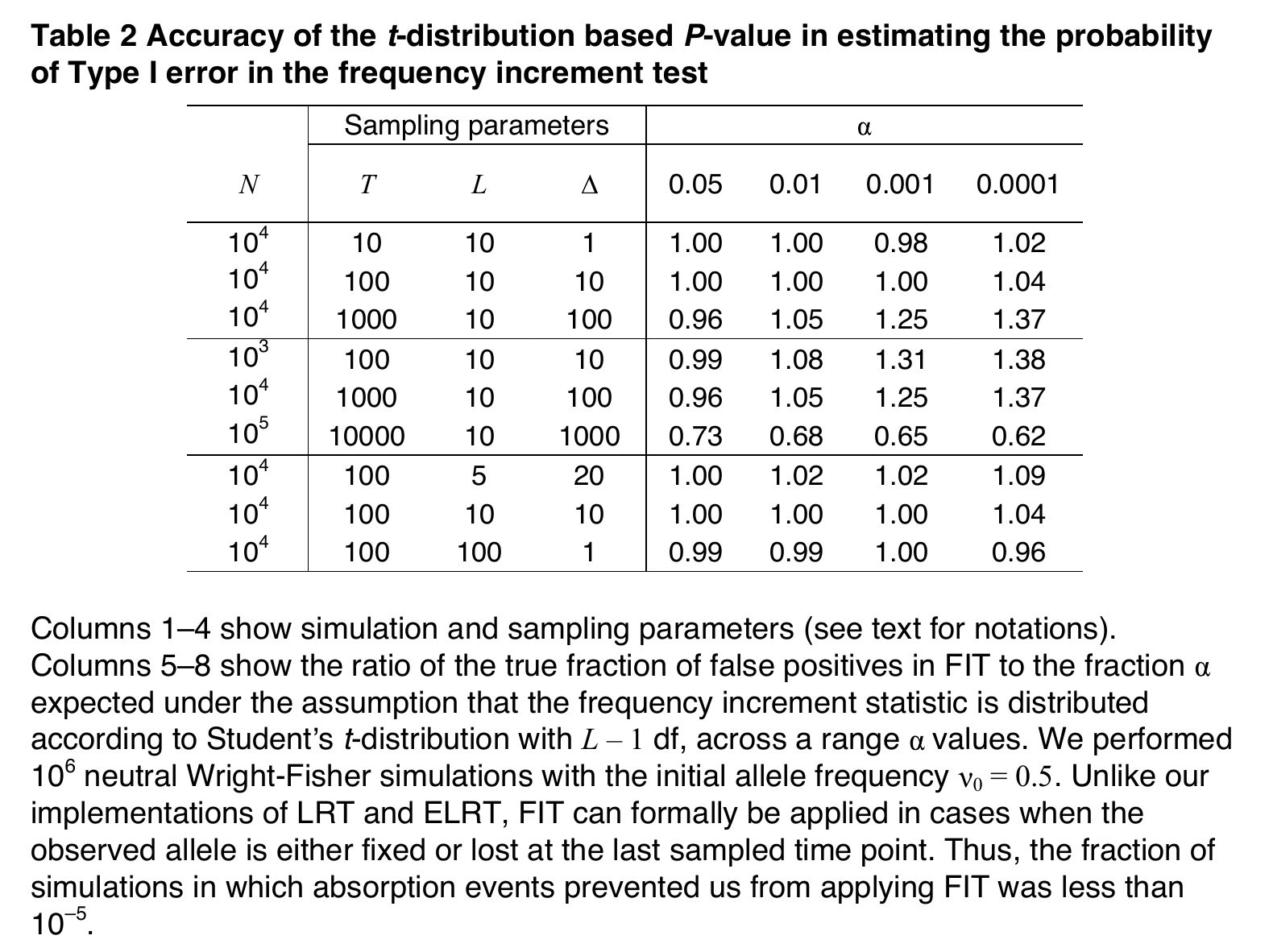}
\end{table}

\begin{sidewaystable}[th]
\centering
\includegraphics{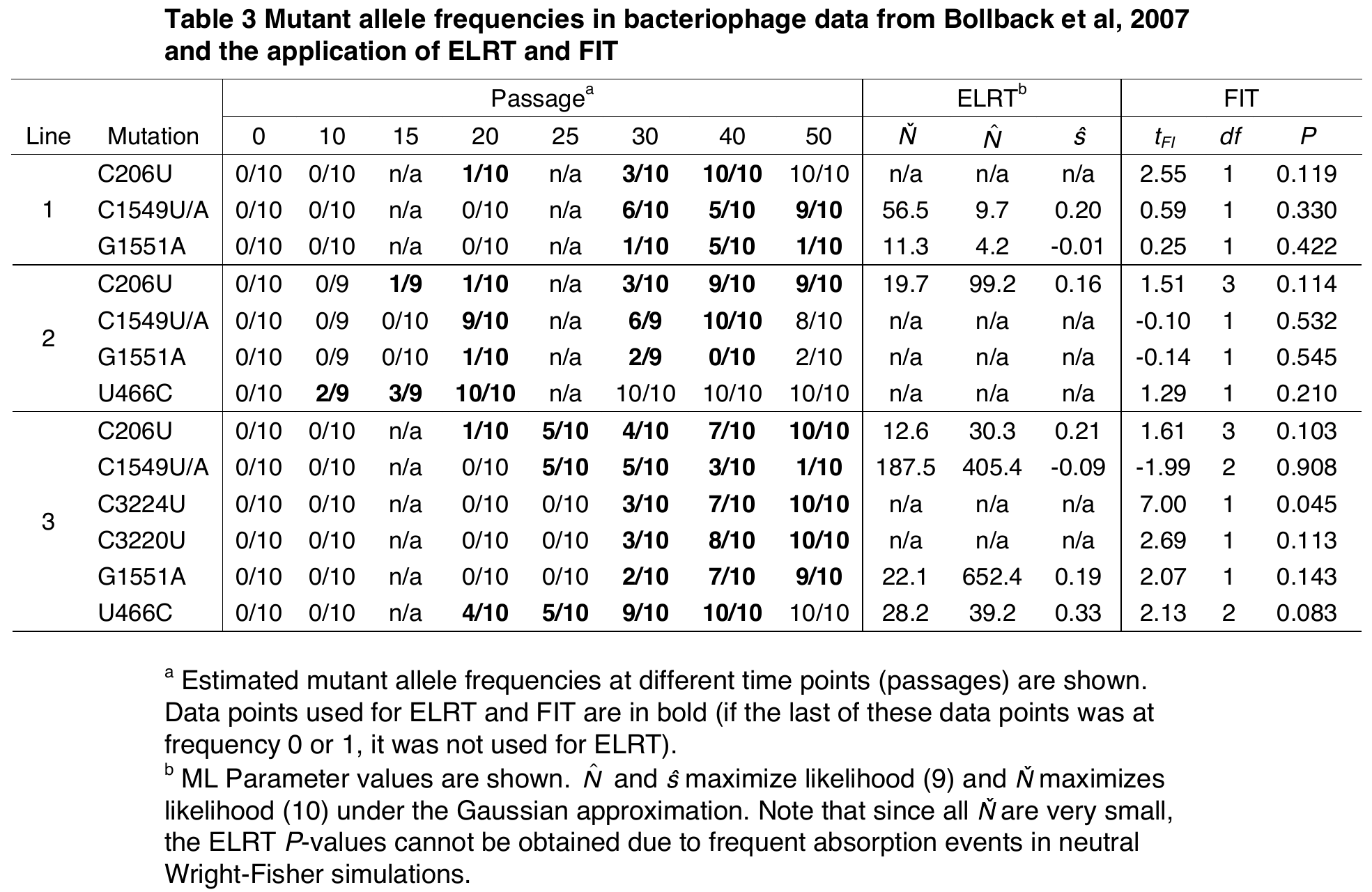}
\end{sidewaystable}

\begin{table}[th]
\centering
\includegraphics{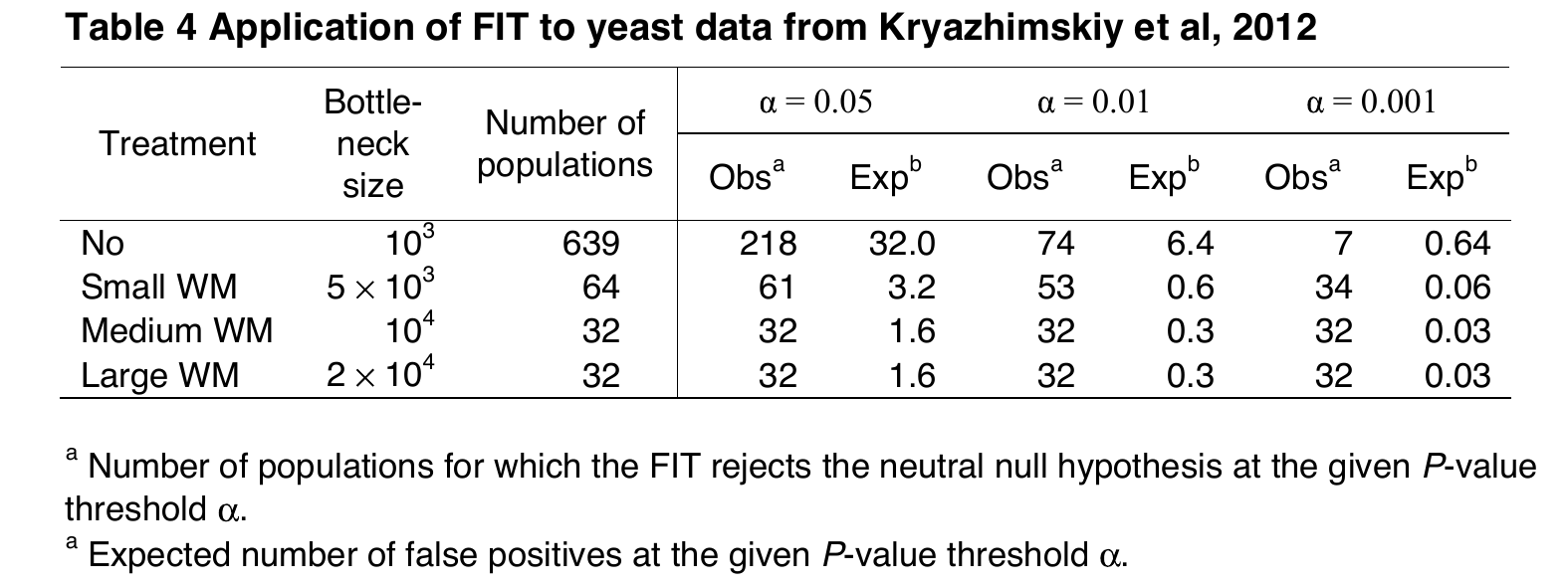}
\end{table}

\clearpage

\section*{Supplementary figures and tables}

\begin{figure}[th]
\begin{center}
\includegraphics{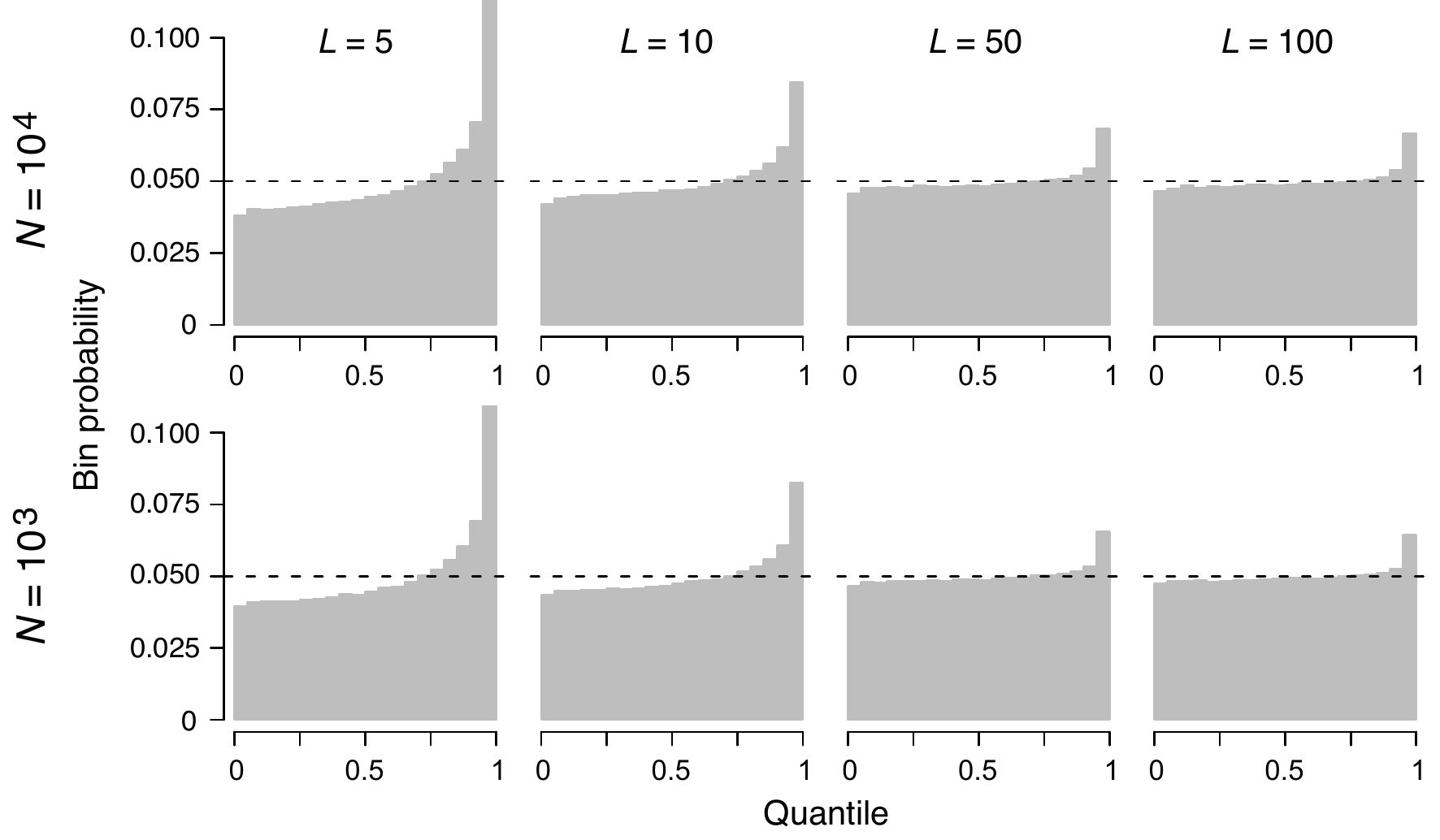}
\end{center}
\caption*{
  {\bf Figure S1. Comparison of the true LRS distribution to the $\chi^2$ distribution with 1 df.}
  Panels show comparisons for different values of the population size $N$ and the number of sampled time points $L$, as indicated on the left and on top.
  Notations are as in Figure~\ref{fig:quantiles}.
  Parameter values: $T = 100$, $\Delta=20$,  $\nu_0 = 0.5$; the number of Wright-Fisher simulations was $10^6$.
}
\label{suppfig:quantiles}
\end{figure}

%[AF:EDIT] added 'the' to 'the number of Wright-Fisher simulations...'

%\clearpage

\begin{figure}[th]
\centering
\includegraphics{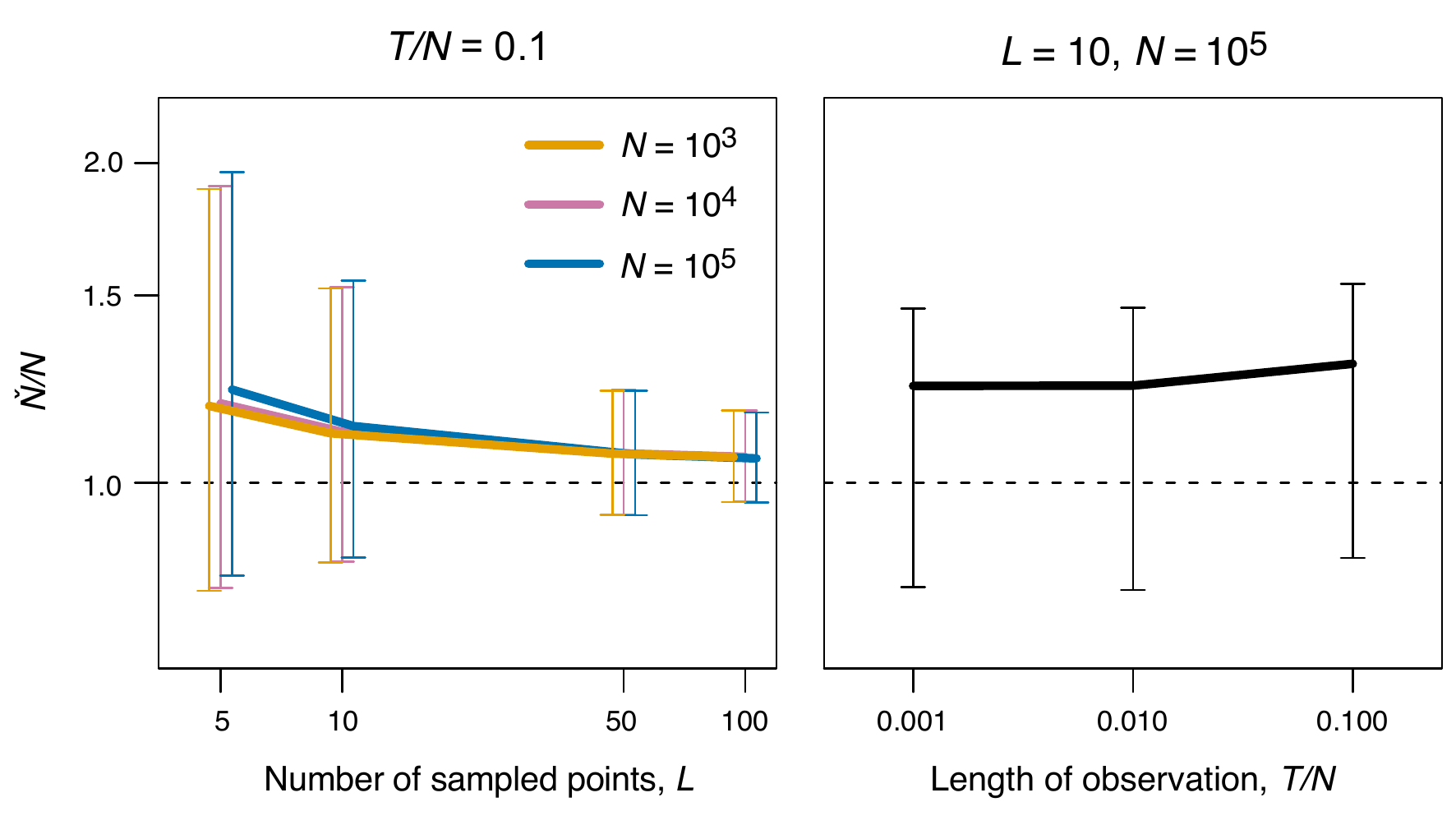}
\caption*{
  {\bf Figure S2. Bias in the maximum-likelihood estimate of population size under neutrality. }
  The figure shows the ratio of the most-likely population size under neutrality,
  $\check N$, to the true population size, $N$, as a function of the number of sampled points $L$ (left panel) 
  and as a function of the length of the observed time series $T$ (right panel).
  Whiskers indicate quartiles of the distribution of $\check N/N$.
  In the right panel, curves for different population sizes are slightly shifted along the $x$-axis for clarity.
  Bias in $\check N$ decreases as the number of sampled time points increases.
  The bias is nearly independent of $N$ and of the length of the sampling period.
 The number of Wright-Fisher simulations was $10^5$.
}
\label{suppfig:Nbias}
\end{figure}

%\clearpage

\begin{figure}[th]
\centering
\includegraphics{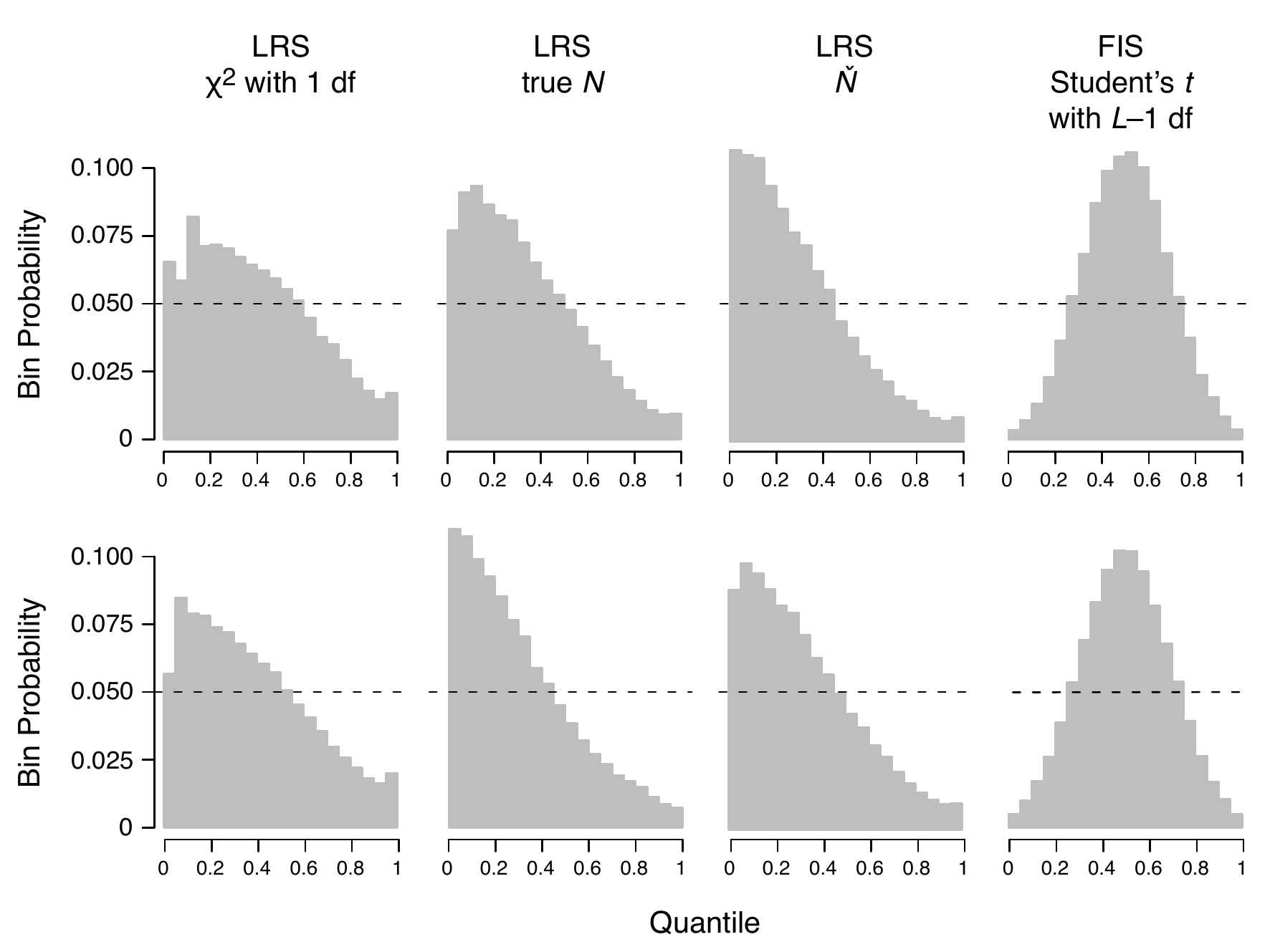}
\caption*{
  {\bf Figure S3. Distributions of various test statistics under the neutral null
  hypothesis, when allele frequencies are sampled with noise.}
  Top row, $n = 50$.
  Bottom row, $n = 100$.
  Notations as in Figure~\ref{fig:quantiles}.
  Parameter values:  $N = 10^3$, $T = 10$, $\Delta=2$,  $L=5$, $\nu_0 = 0.5$; the number of Wright-Fisher simulations was $10^5$.

}
\label{suppfig:quantiles sampled}
\end{figure}

%[AF:EDIT] added 'the' to 'the number of Wright-Fisher simulations...'

\clearpage

\begin{table}[th]
\centering
\includegraphics{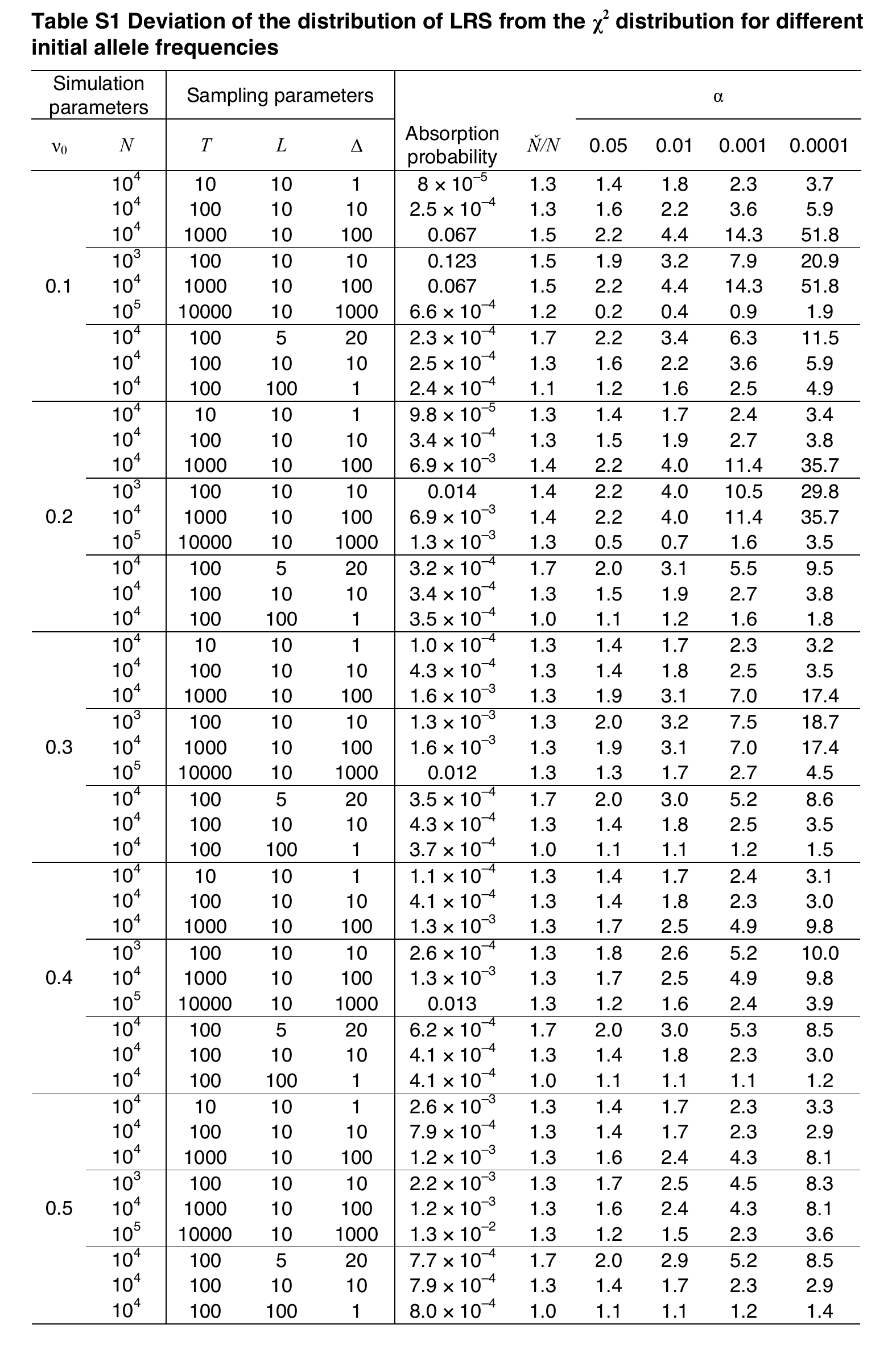}
\end{table}

\begin{sidewaystable}[th]
\centering
\includegraphics{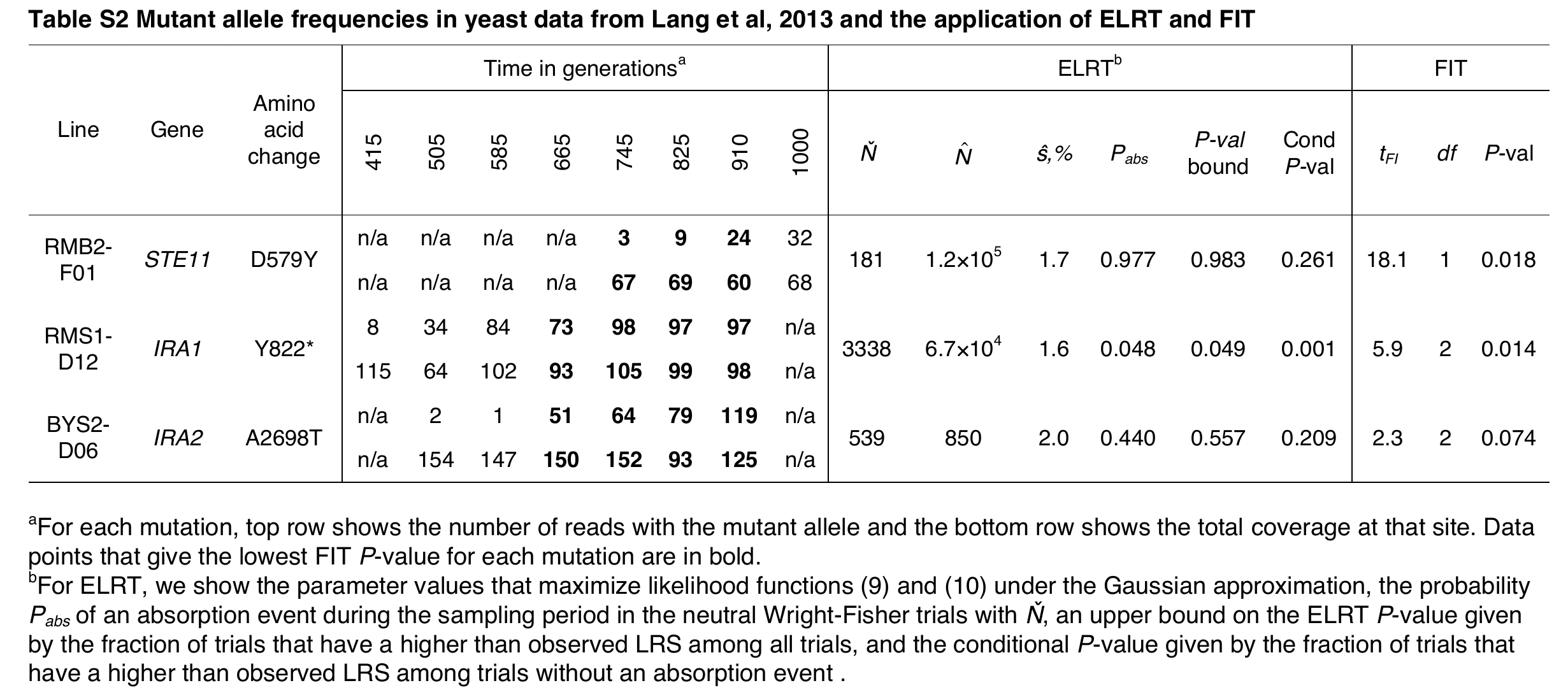}
\end{sidewaystable}

\end{document}